\documentstyle[eqsecnum,prd,aps]{revtex}
\tighten
\begin{document}

\draft

\title{Relativistic second-order perturbations of the Einstein-de Sitter
Universe} 

\author{Sabino Matarrese,$^{1}$ Silvia Mollerach$^{2}$ and Marco Bruni$^{3}$ }
\address{$^{1}$Dipartimento di Fisica ``G. Galilei'', Universit\`a di
Padova, via Marzolo 8, 35131 Padova, Italy} 
\address{$^{2}$Departamento de Astronomia y Astrof\'\i sica,
 Universidad de Valencia, 46100 Burjassot, Valencia, Spain}
\address{$^{3}$SISSA -- International School for Advanced Studies,
Via Beirut 2--4, 34014 Trieste, Italy}
\date{July 25, 1997}

\maketitle
\thispagestyle{empty}

\begin{abstract}

We consider the evolution of relativistic perturbations in the
Einstein-de Sitter cosmological model, including second-order
effects. The perturbations are considered in two different settings:
the widely used synchronous gauge and the Poisson (generalized
longitudinal) one. Since, in general, perturbations are gauge dependent,
we start by considering gauge transformations at second order. Next,
we give the evolution of perturbations in the synchronous gauge,
taking into account both scalar and tensor modes in the initial
conditions. Using the second-order gauge transformation previously
defined, we are then able to transform these perturbations to the
Poisson gauge.  The most important feature of second-order
perturbation theory is mode-mixing, which here also means, for
instance, that primordial density perturbations act as a source for
gravitational waves, while primordial gravitational waves give rise to
second-order density fluctuations.  Possible applications of our
formalism range from the study of the evolution of perturbations in
the mildly non-linear regime to the analysis of secondary anisotropies
of the Cosmic Microwave Background.

\end{abstract}


\newpage


\section{Introduction}

\setcounter{equation}{0}


The study of the evolution of cosmological perturbations is of primary
importance for understanding the present properties of the large-scale
structure of the Universe and its origin. This study is usually
performed with different techniques and approximations, depending on
the specific range of scales under analysis.  So, for scales 
well within the Hubble radius, the analysis of the 
gravitational instability of collisionless matter 
is usually restricted to the Newtonian approximation. As
seen in the Eulerian picture, this approximation basically consists in
adding a first-order lapse perturbation $2\varphi_g/c^2$ to the line
element of a matter--dominated Friedmann--Robertson--Walker (FRW)
model, while keeping non--linear density and velocity perturbations
around the background solution. The peculiar gravitational potential
$\varphi_g$ is determined by the dimensionless matter-density contrast
$\delta$ via the cosmological Poisson equation, $\nabla^2 \varphi_g = 4
\pi G a^2 \varrho_b \delta$, with $\varrho_b$ the background matter
density and $a$ the FRW scale--factor. The fluid dynamics is then
studied by accounting for mass and momentum conservation, to close the
system (see, e.g., Ref. \cite{bi:pjep80}).  This procedure is thought
to produce accurate results on scales much larger than the
Schwarzschild radius of collapsing bodies but much smaller than the
Hubble horizon, where $\varphi_g/c^2$ keeps much less than unity,
while the peculiar matter flows never become relativistic. The
first-order matter perturbations obtained with this Newtonian treatment 
can be shown to coincide with the results of linear general relativistic
perturbation theory in the so-called longitudinal gauge\cite{bi:bertschinger}. 
To second order, however, the comparison is made
non-trivial by the occurrence of non-linear post-Newtonian (and higher
order in $1/c$) terms in the relativistic theory (see also
Refs.\cite{bi:MPS,bi:MT}). Some, but not all, of the aspects of the
relativistic treatment can be accounted for by adding an extra
post-Newtonian perturbation $-2\varphi_g/c^2$ to the conformal spatial
metric, an extension that leads to the so-called weak-field approximation (see,
e.g. Ref.  \cite{bi:pjep93}). This improvement allows, for instance, a
rather accurate treatment of photon trajectories in the geometry
produced by matter inhomogeneities, as required in the
study of gravitational lensing by cosmic structures (see, e.g.,
Ref. \cite{bi:SEF}) and other applications. 
It is worth mentioning that the full
post-Newtonian line-element in Eulerian coordinates would also include
non-vanishing shift components (see, e.g., Ref.\cite{bi:KP95}).  A
second-order perturbative expansion starting from this metric would
lead to the same result of our Poisson-gauge approach discussed below,
with the obvious exception of those terms which are
post-post-Newtonian or higher in a $1/c$ expansion.  

>From the point of view of the Lagrangian frame of the matter, corresponding 
to our synchronous and comoving gauge below, the Newtonian approach is quite
different: the `Newtonian Lagrangian metric' can be cast in a simple
form, where the spatial metric tensor is written in terms of the
Jacobian matrix connecting Lagrangian to Eulerian coordinates.
According to this approach, post-Newtonian terms of any order appear
as spatial metric perturbations over this `Newtonian
background'\cite{bi:MT}.  Without discussing the long list of
cosmological approximation schemes which have been proposed to follow
the non-linear dynamics of collisionless matter in the Newtonian
framework, let us only mention the celebrated Zel'dovich
approximation\cite{bi:ZEL}, which is strictly related to the
Lagrangian Newtonian approach. Various extensions of Zel'dovich theory
to the relativistic case have been proposed in the literature; all of
them however require either global or local symmetries, thereby
preventing their application to the cosmological structure formation
problem.  A relativistic formulation of the Zel'dovich approximation,
assuming no limitations on the initial conditions, is instead
introduced in Ref.\cite{bi:MD97}.

So far the list basically covers all those methods which have been
devised to follow non-linear structure formation by gravitational
instability in the Universe, with the only possible exception of a few
relevant exact solutions of Einstein's field equations, such as the
Tolman-Bondi one and some of the Bianchi and Szekeres solutions (see,
e.g., Ref.\cite{bi:BMP} and references therein for a review). These
exact solutions, however, have only limited application to realistic
cosmological problems.

The study of small perturbations giving rise to large-scale
temperature anisotropies of the Cosmic Microwave Background is instead
usually treated with the full technology of first-order relativistic
perturbation theory, either in a gauge-invariant fashion or by
specifying a suitable gauge. On small and intermediate angular scales, however,
where the description in terms of first-order perturbation theory is
no longer accurate, second-order metric perturbations can play a
non-trivial role and determine new contributions, such as those
leading to the non-linear Rees-Sciama effect \cite{bi:RS68}. In such a
case a fully general relativistic treatment is needed, such as that
recently put forward by Pyne and Carroll\cite{bi:pyne} and implemented
in second-order perturbation theory by Mollerach and
Matarrese\cite{bi:mm97}.

The aim of the present paper is to provide a complete account of
second-order cosmological perturbations in two gauges: the synchronous
and comoving gauge and the so-called Poisson one, a generalization of
the longitudinal gauge discussed by
Bertschinger\cite{bi:bertschinger} and Ma and
Bertschinger\cite{bi:mabe}. The former was chosen here because of the
advantages it presents in performing perturbative calculations.  The
latter, on the other hand, being the closest to the Eulerian Newtonian
picture, allows a simpler physical understanding of the various
perturbation modes.  The link between these gauges is provided by a
second-order gauge transformation of all the geometrical and physical
variables of the problem.  The general problem of non-linear gauge
transformations in a given background spacetime has been recently
dealt with by Bruni et al.\cite{bi:P1} (see also \cite{bi:P2}), 
and will be shortly reviewed in the following section.

The range of applicability of our general relativistic second-order
perturbative technique is that of small fluctuations around a FRW
background, but with no extra limitations.  
It basically allows to describe perturbations down to scales which 
experience slight departures from a linear behavior, which, in present-day 
units, would include all scales above about $10$ Mpc in any realistic scenario 
of structure formation.  Accounting for second-order effects generally
helps to follow the gravitational instability on a longer time-scale
and to include new non-linear and non-local phenomena. The advantage
of such a treatment is precisely that it enables one to treat a large
variety of phenomena and scales within the same computational
technique.

The literature on relativistic second-order perturbation theory in a
cosmological framework is not so vast. The pioneering work in this
field is by Tomita\cite{bi:tomita67}, who, back in 1967, performed a
synchronous-gauge calculation of the second-order terms produced by
the mildly non-linear evolution of scalar perturbations in the
Einstein--de Sitter universe. Matarrese, Pantano and S\'aez\cite{bi:MPS}
obtained an equivalent result, but with a different technique, in
comoving and synchronous coordinates. Using a tetrad formalism, Russ
et al.\cite{bi:russ} recently extended these calculations to include
the second-order terms generated by the mixing of growing and decaying
linear scalar modes.  Salopek, Stewart and Croudace\cite{bi:ssc}
applied a gradient expansion technique to the calculation of
second-order metric perturbations. The inclusion of vector and tensor
modes at the linear level, acting as further seeds for the origin of
second-order perturbations of any kind (scalar, vector and tensor),
has been, once again, first considered by Tomita
\cite{bi:tomita71-72}.

In this paper we study the second-order perturbations of an irrotational 
collisionless fluid in the Einstein-de Sitter background, including both 
growing-mode scalar perturbations and gravitational waves at the linear level.
The plan of the paper is as follows. In the next section we
summarize the results of Ref.\cite{bi:P1} regarding non--linear gauge
transformations for perturbations of any given background spacetime. 
In Section III we consider perturbations of a generic flat Robertson--Walker 
model, and we give the transformations between any 
two given  gauges, up to second order. Section IV
is devoted to the study of the evolution of perturbations in the
synchronous gauge, in the specific case of irrotational dust in the Einstein 
de-Sitter background. In Section V we apply the formulas obtained in Section 
III to obtain  the transformations between the synchronous\cite{bi:li46} and 
the Poisson (generalized longitudinal\cite{bi:bertschinger}) 
gauge. Using these transformations, in Section VI the results  of Section IV 
are transformed to the Poisson gauge. Section VII contains a final discussion.
Appendix A reviews some mathematical results obtained in Ref.\cite{bi:P1}
and used in Section II; Appendices B and C contain useful formulas in
the synchronous gauge, used in Section IV. Appendix D contains formulas used 
to obtain some of the Poisson-gauge results in Section VI. 


\section{Gauge dependence at second and higher order}
\label{sec:GT}
\setcounter{equation}{0}


The idea underlying the theory of spacetime perturbations is the same
that we have in any perturbative formalism: we try to find approximate
solutions of some field equations, regarding them as `small'
deviations from a known exact background solution. The basic difference
arising in general relativity, or in other spacetime theories, is that
we have to deal with perturbations not only of fields in a given
geometry, but of the geometry itself.

The perturbation $\Delta T$ in any relevant quantity, say represented
by a tensor field $T$, is defined as the difference between the value
$T$ has in the physical spacetime (the perturbed one), and the value
$T_0$ the same quantity has in the given (unperturbed) background
spacetime.  However, it is a basic fact of differential geometry that,
in order to make the comparison of tensors meaningful at all, one has
to consider them at the same point.  Since $T$ and $T_0$ are defined
in different spacetimes, they can thus be compared only after a
prescription for identifying points of these spacetimes is given.  A
{\em gauge choice\/} is precisely this, i.e., a one-to-one
correspondence (a map) between the background and the physical
spacetime.  A change of this map is then a gauge transformation, and
the freedom one has in choosing it gives rise to an arbitrariness in
the value of the perturbation of $T$ at any given spacetime point,
unless $T$ is gauge-invariant.  This is the essence of the `gauge
problem', which has been discussed -- mainly in connection with
linear perturbations -- in many
papers\cite{bi:sachs,bi:SW,bi:JB,bi:EB,bi:stewart} and review
articles\cite{bi:KS,bi:MFB}, following different approaches.

In order to discuss in depth higher-order perturbations and gauge
transformations, and to define gauge invariance, one needs to formalise 
the above ideas, giving a precise geometrical description of what
perturbations and gauge choices are.  In a previous paper\cite{bi:P1}
(see also \cite{bi:P2}) we have considered this problem in great
detail, following in the main the approach used in
Refs.\cite{bi:SW,bi:stewart,bi:waldbook}. 
Instead of  directly considering perturbations,  we have
first looked upon the geometry of the problem in  full  generality, taking an
exact (i.e., non perturbative) point of view; after that, we have
expanded all the geometrical quantities in appropriately defined
Taylor series, thus going beyond the usual linear treatment.

In this section we shall summarize the main results obtained in
 Ref.\cite{bi:P1}. Appendix A explains in some  more
detail how to proceed in the calculations. Here and in the following 
 Greek indices
$\mu,\nu,\ldots$ take values from 0 to 3, and the Latin ones
$i,j,\ldots$ from 1 to 3; units are such that $c=1$.

We finally remind here  some basics 
about the Lie derivative along a vector field $\xi$, which will be
useful in the following.
The Lie derivative of any tensor $T$  
 of type $(p,q)$ (a tensor with $p$
 contravariant and $q$ covariant indices, that we omit here and in the
 following) is also a tensor of the same type $(p,q)$.
For a  scalar $f$, a contravariant vector $Z$ and a covariant tensor
$T$ of rank two, the expressions of the Lie derivative along $\xi$ 
are, respectively:
\begin{eqnarray}\label{eq:liedev1}
\pounds_\xi f & = &  f_{,\mu}\xi^\mu\; ; 
\\
\label{eq:liedev2}
\pounds_\xi Z^\mu & = & Z^\mu_{,\nu}\xi^\nu -\xi^\mu_{,\nu}Z^\nu\; ;
\\
\label{eq:liedev3}
\pounds_\xi T_{\mu\nu} & = & T_{\mu\nu,\sigma}\xi^\sigma
+\xi^\sigma_{,\mu} T_{\sigma\nu}+\xi^\sigma_{,\nu} T_{\mu\sigma}\; .
\end{eqnarray}
Expressions for any other tensor can easily be derived from these.
A second or higher Lie derivative is easily defined from these
formulas; e.g., for a vector we have $\pounds^2_\xi
Z=\pounds_\xi(\pounds_\xi Z)$: since one clearly sees from
(\ref{eq:liedev2}) that $\pounds_\xi Z$ is itself a contravariant
vector, one needs only to apply  (\ref{eq:liedev2}) two times to obtain the
components of $\pounds^2_\xi Z$. Similarly, one derives expressions
for the second Lie derivative of any tensor.

\subsection{Gauge transformations: an exact point of view}

A basic assumption in perturbation theory is the existence of a
parametric family of solutions of the field equations, to which the
unperturbed background spacetime belongs\cite{bi:waldbook}.  In
cosmology and in many other cases in general relativity, one deals
with a one-parameter family of models ${\cal M}_\lambda$; $\lambda$ is
real, and $\lambda=0$ identifies the background ${\cal M}_0$.  On each
${\cal M}_\lambda$ there are tensor fields $T_\lambda$ representing
the physical and geometrical quantities (e.g., the metric).  The
parameter $\lambda$ is used for Taylor expanding these $T_\lambda$; 
the physical spacetime
${\cal M}_\lambda$ can eventually be identified by $\lambda=1$. 
The aim of perturbation theory is to construct an approximated solution
 to ${\cal M}_\lambda$.

Each one-to-one correspondence between points of ${\cal M}_0$ and
points of ${\cal M}_\lambda$ is thus a one-parameter function of
$\lambda$: we can represent two such `point identification
maps'\cite{bi:SW} as $\psi_\lambda$ and $\varphi_\lambda$ (for a
depiction of this and the following, see Fig. 1 and 2 in
\cite{bi:P1}).  Suppose that coordinates $x^\mu$ have been assigned on
the background ${\cal M}_0$, labeling the different points.  A
one-to-one correspondence, e.g. $\psi_\lambda$, carries these
coordinates over ${\cal M}_\lambda$, and defines a choice of gauge:
therefore, it is natural to call the correspondence itself `a gauge'.
A change in this correspondence, keeping the background coordinates
fixed, is a gauge transformation\cite{bi:JB}.  Thus, let $p$ be any
point in ${\cal M}_0$, with coordinates $x^\mu(p)$, and let us use the
gauge $\psi_\lambda$: $O=\psi_\lambda(p)$ is the point in ${\cal
M}_\lambda$ corresponding to $p$, to which $\psi_\lambda$ assigns the
same coordinate labels. However, we could as well use a different
gauge, $\varphi_\lambda$, and think of $O$ as the point of ${\cal
M}_\lambda$ corresponding to {\it a different point} $q$ in the
background, with coordinates $\tilde{x}^\mu$: then $O=\psi_\lambda(p)=
\varphi_\lambda(q)$.  Thus, the change of the correspondence, i.e. the
gauge transformation, may actually be seen as 
 a {\it one-to-one correspondence
between different points in the background}. Since we start from a
point $p$ in ${\cal M}_0$, we carry it over to $O=\psi_\lambda(p)$ in
${\cal M}_\lambda$, and then we may come back to $q$ in ${\cal M}_0$
with $\varphi_\lambda^{-1}$, i.e. $q=\varphi_\lambda^{-1}(O)$, the
overall gauge transformation is also a function of $\lambda$, which we
may denote as $\Phi_\lambda$, and is given by composing
$\varphi_\lambda^{-1}$ with $\psi_\lambda$, so that we can write
$q=\Phi_\lambda(p):= \varphi_\lambda^{-1}(\psi_\lambda(p))$.  We then
have that the coordinates of $q$, $\tilde{x}^\mu(\lambda,
q)=\Phi^\mu_\lambda(x^\alpha(p))$, are one-parameter functions of
those of $p$, $x^\alpha(p)$.  Such a transformation, that in one given
coordinate system moves each point to another, is often called `an
active coordinate transformation', as opposed to passive ones, that
change coordinate labels to each point (see Appendix A).

Now, consider the tensor fields $T_\lambda$ on each ${\cal
 M}_\lambda$.  With the gauges $\varphi_\lambda$ and $\psi_\lambda$ we
 can define, in two different manners, a representation on ${\cal
 M}_0$ of each $T_\lambda$: we can denote these simply by $T(\lambda)$
 and $\tilde{T}(\lambda)$, respectively. These are tensor fields
 defined on ${\cal M}_0$ in such a way that each of them has, in the
 related gauge, the same components of $T_\lambda$. On the other hand,
 $T(\lambda)$ and $\tilde{T}(\lambda)$ are related by $\Phi_\lambda$,
 which gives rise to a relation between their components given by Eq.\
 (\ref{eq:Tpb1}).  Since in each gauge we now have a field
 representing $T_\lambda$ on ${\cal M}_0$, at each point of the
 background we can compare these fields with $T_0$, and define
 perturbations.  In the first gauge the total perturbation is $\Delta
 T (\lambda):=T(\lambda)-T_0$, and in the second one is $\Delta
 \tilde{T} (\lambda):=\tilde{T}(\lambda)-T_0$.  This non--uniqueness
 is the gauge dependence of the perturbations.

It should be noted at this point that we haven't so far made any
approximation: the definitions given above are exact, and hold in
general, to any perturbative order. 

\subsection{Gauge transformations: second order expansion}

In order to proceed and compute at the desired order of accuracy in
$\lambda$ the effects of a gauge transformation, we need Taylor
expansions. In this respect, a crucial point is that gauge choices
such as $\psi_\lambda$ and $\varphi_\lambda$ form one-parameters groups
with respect to $\lambda$, while  the gauge transformations
$\Phi_\lambda$ form a one-parameter
family that in general is not a group (see Appendix A and \cite{bi:P1,bi:P2}
for more details). 
Only in linear theory the
action of $\Phi_\lambda$ is approximated by that of the element of 
a one-parameter group of transformations.
A one-parameter group of transformations is associated with a vector
field $\xi$ and the congruence it generates, and therefore at first order in
$\lambda$ the effect of $\Phi_\lambda$ on the coordinates $x^\mu(p)$
is approximated by $\tilde{x}^\mu\simeq x^\mu +\lambda\xi^\mu$,
whereas for a tensor $T$ we have $\tilde{T}\simeq T+\lambda
\pounds_\xi T$,    as it
is well known. However,
the fact that $\Phi_\lambda$ does not form a group comes into play
with non linearity,  
and  one can show that at second order two vector fields $\xi_{(1)}$ and
$\xi_{(2)}$ are involved, and so on. That is,  
 the Taylor expansion of a one-parameter
family of transformations  $\Phi_\lambda$ involves, at a given order
$n$, $n$ vector fields $\xi_{(k)}$, $k=1\ldots n$. 
At second order, the expansion of the transformation
$\tilde{x}^\mu(\lambda)=\Phi^\mu_\lambda(x^\alpha)$  between the coordinates
of any pair of points of the background mapped into one another by
$\Phi_\lambda$ gives
\begin{equation}
 \tilde{x}^\mu(\lambda)=x^\mu
+\lambda\,\xi_{(1)}^\mu+{\lambda^2\over
2}\,\left({\xi_{(1)}^\mu}_{,\nu}\xi_{(1)}^\nu
+\xi_{(2)}^\mu\right)+{\cal O}(\lambda^3)\;.
\label{eq:GT1}
\end{equation}
This is often called an 
`infinitesimal point transformation'. From this, one can always define
(again, see Appendix A) an associated 
ordinary (passive) coordinate transformation,
Eq. (\ref{lemma2coord''}).
Substitution of this into Eq. (\ref{eq:Tpb1})  gives -- after properly
collecting terms --
the gauge transformation for a  tensor $T$: 
\begin{equation}
\tilde{T}(\lambda)=T(\lambda)+\lambda \pounds_{\xi_{(1)}} T
+\frac{\lambda^2}{2}\left(\pounds^2_{\xi_{(1)}}
+\pounds_{\xi_{(2)}}\right)T + {\cal O}(\lambda^3) \;.
\label{eq:GT2}
\end{equation}
If $T$ is a tensor of type $(p,q)$, the components of 
each term in this formula have $p$ 
 contravariant and $q$ covariant indices, appropriately given by the
rules (\ref{eq:liedev1})--(\ref{eq:liedev3}).

A simple heuristic argument that may  help understanding why two vector
fields are involved in the second order gauge transformations (and $n$
vectors at $n$-th order) is the following. In practice, we usually
consider the gauge transformation between two given gauges, e.g., in
this paper, the
synchronous and the Poisson ones. Therefore, having the conditions
that fix the two gauges, the unknowns of the
problem are the degrees of freedom that allow us to pass from one gauge
to another. Then, consider the usual first order gauge transformation,
i.e., the first order part of (\ref{eq:GT1}), (\ref{eq:GT2}): it is
clear that this fully determines $\xi_{(1)}$ as a field in the
background. Going to second order, it should be clear   by the very 
 definition of a Taylor expansion that 
$\xi_{(1)}$ itself cannot depend on $\lambda$, and its contribution
to second order is built from what we already know at first order,
i.e. it can only give a quadratic contribution. On the other hand,
since at second order in both gauges
 we have new degrees of freedom in any quantity (with respect to first
order), it is clear that the
gauge transformation itself must contain new degrees of freedom: these
are given by $\xi_{(2)}$.

Since in the two gauges we can write, respectively\cite{bi:defpert}:
\begin{eqnarray}
T(\lambda)&=&T_0 +\lambda\delta T +\frac{\lambda^2}{2}\delta^2 T +{\cal
O}(\lambda^3)\;,
\label{eq:gc1}
\\
\tilde{T}(\lambda) &=& T_0 +\lambda\delta \tilde{T} 
+\frac{\lambda^2}{2}\delta^2 \tilde{T} +{\cal O}(\lambda^3)\;,
\label{eq:gc2}
\end{eqnarray}
we can substitute (\ref{eq:gc1}) and (\ref{eq:gc2}) into
(\ref{eq:GT2}) in order to obtain, at first and second order in
$\lambda$, the required gauge transformations for $\delta T$ and
$\delta^2 T$:
\begin{eqnarray}
\label{eq:t1}
\delta \tilde{T} &=&\delta T +\pounds_{\xi_{(1)}} T_{0}\, ,\\
\label{eq:t2} 
\delta^2 \tilde{T}& =&\delta^2 T +2\pounds_{\xi_{(1)}} \delta T
+\pounds^{2}_{\xi_{(1)}} T_{0} +\pounds_{\xi_{(2)}}
T_{0}\;.
\end{eqnarray}
Eq. (\ref{eq:t1}) is the well known result mentioned
above. Eq. (\ref{eq:t2}) gives  the general gauge
transformation for second order perturbations, and shows that this 
 is made up of three parts: the first couples the first
order generator of the transformation $\xi_{(1)}$ with the first order
perturbation $\delta T$; the second part couples the zero order
background $T_0$ with terms quadratic in $\xi_{(1)}$; the last part
couples $T_0$ with the second order generator $\xi_{(2)}$ of the
transformation, in the same manner than the term $\pounds_{\xi_{(1)}}
T_0$ does at first order, in (\ref{eq:t1}).
Eq. (\ref{eq:t2}) also shows that there are special second order  gauge
transformations, purely due to $\xi_{(2)}$, when $\xi_{(1)}=0$; on the
other hand, a non vanishing $\xi_{(1)}$ always affects second order
perturbations (cf. \cite{bi:gleiseretal}). Finally, in some specific
problems in which only effects quadratic in first order perturbations
are important, one can consider $\xi_{(2)}=0$: for instance, this is
the case of back-reaction effects (cf. Ref. \cite{bi:MAB}). 

\subsection{Gauge invariance}

It is logically possible to establish a condition for gauge invariance
to a given perturbative order $n$ even without knowledge of the gauge
transformation rules holding at that order; see \cite{bi:P1}. However,
we shall focus here on gauge invariance to second order, using
Eqs. (\ref{eq:t1}), (\ref{eq:t2}).

We need to state a clear definition of gauge invariance before
giving a condition. The most natural definition 
is that a tensor $T$  is gauge--invariant to
order $n$ if and only if $\delta^k \tilde{T}=\delta^k T$  for
every $k\leq n$ (we define $\delta^0 T:=T_0$, $\delta T:=\delta^1 T$). 
Thus,  a tensor $T$ is gauge--invariant to second order if
$\delta^2\tilde{T}=\delta^2 T$ {\it and}
$\delta\tilde{T}=\delta T$.
Then, from (\ref{eq:t1}) and (\ref{eq:t2}) we see that, since
$\xi_{(1)}$ and $\xi_{(2)}$ are arbitrary, this condition implies that
$\pounds_\xi T_0=0$ and $\pounds_\xi \delta T =0$, for every vector
field $\xi$ in the background ${\cal M}_0$.
Therefore, apart from trivial cases -- i.e., constant scalars and 
combinations of Kroneker deltas with constant coefficients -- 
gauge invariance to second order requires that $T_0=0$ and
$\delta T=0$ in any gauge. This condition generalizes to second order
the results for first order gauge--invariance that can be found in the
literature, and is easily extended to order $n$; see Ref. \cite{bi:P1}.


\section{Gauge transformation in a flat cosmology up to second order}


In view of the application that will follow in Section V, we shall now
use the formulas obtained in the previous section to 
 show how the perturbations on a spatially flat
Robertson--Walker background in two different gauges are related, up
to second order. This will also introduce the notation used in all the
following sections. Here and in the following 
Latin indices are raised and lowered  
using $\delta^{ij}$ and $\delta_{ij}$, respectively. As discussed before,
we set $\lambda=1$ to describe the physical space-time.

\subsection{Perturbed flat Robertson--Walker universe}

 We shall first consider the metric perturbations,
then those in the energy density and 4-velocity of the matter.

The components of a perturbed spatially flat Robertson--Walker 
metric can be written as 
\begin{equation}\label{eq:m1}
g_{00}=-a^2(\tau)\left(1+2\sum_{r=1}^{+\infty}{1\over
r!}\ \psi^{(r)}\right)\;,
\end{equation}
\begin{equation}\label{eq:m2}
g_{0i}=a^2(\tau)\sum_{r=1}^{+\infty}{1\over
r!}\ \omega^{(r)}_i\;,
\end{equation}
\begin{equation}\label{eq:m3}
g_{ij}=a^2(\tau)\left\{\left[1-2\left(\sum_{r=1}^{+\infty}{1\over
r!}\ \phi^{(r)}\right)\right]\delta_{ij}+\sum_{r=1}^{+\infty}{1\over
r!}\ \chi^{(r)}_{ij}\right\}\;,
\end{equation}
where $\chi^{(r)i}_{i}=0$, and $\tau$ is
the conformal time.  The functions $\psi^{(r)}$, $\omega^{(r)}_i$,
$\phi^{(r)}$, and $\chi^{(r)}_{ij}$ represent the
 $r$-th order perturbation of the metric.

It is standard to use a non-local splitting of perturbations into the
so-called scalar, vector and tensor parts, where scalar (or
longitudinal) parts are those related to a scalar potential, vector
parts are those related to transverse (divergence-free, or solenoidal)
vector fields, and tensor parts to transverse trace-free tensors.
In our case, the shift $\omega_i^{(r)}$ can be decomposed as
\begin{equation}
\omega_i^{(r)}= \partial_i\omega^{(r)\|} +\omega_i^{(r)\perp}\;,
\end{equation}
where $\omega_i^{(r)\perp}$ is a solenoidal vector, i.e.,
$\partial^i\omega_i^{(r)\perp}=0$.  
Similarly, the traceless part of the
spatial metric can be decomposed at any order as
\begin{equation}
\chi^{(r)}_{ij}= {\rm D}_{ij}\chi^{(r)\|}+\partial_i\chi^{(r)\bot}_j+
\partial_j\chi^{(r)\bot}_i+\chi^{(r)\top}_{ij}\;,
\end{equation}
where $\chi^{(r)\|}$ is a suitable function, $\chi^{(r)\bot}_i$ is a
solenoidal vector field, and 
$\partial^i\chi^{(r)\top}_{ij}=0$; hereafter,
\begin{equation}
{\rm D}_{ij}:=\partial_i\partial_j-
{1\over 3}\,\delta_{ij}\nabla^2\;.
\end{equation}

Now, consider the energy density $\varrho$, or any other scalar
 that depends only  on $\tau$ at zero order:
this can be written as
\begin{equation}\label{eq:mu}
\varrho=\varrho_{(0)}+\sum_{r=1}^{+\infty}{1\over
r!}\ \delta^r \varrho\; .
\end{equation}
For the 4-velocity $u^\mu$ of matter we can write
\begin{equation}\label{eq:4v}
u^\mu=\frac{1}{a}\left(\delta^\mu_0 +\sum_{r=1}^{+\infty}{1\over
r!}\ v^\mu_{(r)}\right)\; .
\end{equation}
In addition, $u^\mu$ is subject to the normalization condition $u^\mu
 u^\nu g_{\mu\nu}=-1$; therefore at any order the time component $v^0_{(r)}$ is
 related to the lapse perturbation, $\psi_{(r)}$.  For the first and
 second-order perturbations we obtain, in any gauge:
\begin{eqnarray}
\label{eq:v0psi1}
v^0_{(1)} &= &-\psi_{(1)}\; ; \\
\label{eq:v0psi2}
v^0_{(2)} &= &-\psi_{(2)}+3\psi^2_{(1)}+2\omega^{(1)}_i v^i_{(1)} 
+v^{(1)}_iv_{(1)}^i\; .
\end{eqnarray}
The velocity perturbation $v^i_{(r)}$ can also be split into a scalar
and vector (solenoidal) part:
\begin{equation}
v^i_{(r)}=\partial^i v_{(r)}^{\|} +v^i_{(r)\perp}\;.
\end{equation} 

As we have seen in the last section, the gauge transformation is
determined by the vectors $\xi_{(r)}$.  Splitting their time and space
parts, one can write
\begin{equation}
\xi_{(r)}^0=\alpha^{(r)}\;,
\end{equation}
and
\begin{equation}
\xi_{(r)}^i=\partial^i\beta^{(r)}+d^{(r)i}\;,
\end{equation}
with $\partial_i d^{(r)i}=0$.


\subsection{First-order gauge transformations}

We begin by reviewing briefly some well-known results about 
first-order gauge transformations, as we shall need them in the following.
As in Section \ref{sec:GT}, we simply denote quantities in the new
gauge by a tilde.


>From Eq.\ (\ref{eq:t1}),
 it follows that the first-order perturbations
of the metric transform as
\begin{equation}
\delta\tilde{g}_{\mu\nu}=\delta
g_{\mu\nu}+\pounds_{\xi_{(1)}}g^{(0)}_{\mu\nu}\;,
\end{equation}
where $g^{(0)}_{\mu\nu}$ is the background metric.  Therefore, using
Eq.\ (\ref{eq:liedev3}), we obtain the following transformations for
the first-order quantities  appearing in 
Eqs.\ (\ref{eq:m1})--(\ref{eq:m3}):
\begin{equation}
\tilde{\psi}_{(1)}=\psi_{(1)}+\alpha_{(1)}^{\prime}
+{a'\over a}\,\alpha_{(1)}\;;
\label{eq:pertpsi}
\end{equation}
\begin{equation}
\tilde{\omega}^{(1)}_i=\omega^{(1)}_i-\alpha^{(1)}_{,i}+
\beta^{(1)\prime}_{,i}+d^{(1)\prime}_i\;;
\end{equation}
\begin{equation}
\tilde{\phi}_{(1)}=\phi_{(1)}-
{1\over 3}\,\nabla^2\beta_{(1)}-{a'\over
a}\,\alpha_{(1)}\;;
\end{equation}
\begin{equation}
\tilde{\chi}^{(1)}_{ij}=\chi^{(1)}_{ij}+
2{\rm D}_{ij}\beta^{(1)}+d^{(1)}_{i,j}+d^{(1)}_{j,i}\;;
\label{eq:pertchi}
\end{equation}
where a prime denotes the derivative with respect to $\tau$.

For a scalar $\varrho$, from Eqs.\ (\ref{eq:t1}), (\ref{eq:mu}), and 
(\ref{eq:liedev1}) we have
\begin{equation}\label{eq:mut}
\delta\tilde{\varrho}=\delta\varrho +\varrho_{(0)}^\prime\alpha_{(1)}\; .
\end{equation}
For the 4-velocity $u^\mu$, we have from Eqs.\ (\ref{eq:t1}) 
\begin{equation}
\delta \tilde{u}^\mu=\delta u^\mu 
+\pounds_{\xi_{(1)}}u^\mu_{(0)}\;.
\end{equation} 
Using Eqs.\ (\ref{eq:liedev2}) and (\ref{eq:4v}) this gives:
\begin{eqnarray}
\label{eq:v0}
\tilde{v}^0_{(1)} & = & v^0_{(1)}
-\frac{a'}{a}\alpha_{(1)}-\alpha_{(1)}^\prime\; ; \\
\label{eq:vi}
\tilde{v}^i_{(1)} & = & v^i_{(1)}
-\beta_{(1)}^{\prime,i}-d_{(1)}^{i\prime}\; .
\end{eqnarray}

The 4-velocity is however subject to the constraint  
(\ref{eq:v0psi1}),
therefore Eq.\ (\ref{eq:v0}) reduces to Eq.\ (\ref{eq:pertpsi}).

\subsection{Second-order gauge transformations}

We now extend these well-known transformation rules of linear
metric perturbations to the second-order.

Second-order perturbations of the metric transform, according to Eq.\
(\ref{eq:t2}), as
\begin{equation}
\delta^2\tilde{g}_{\mu\nu}=\delta^2g_{\mu\nu}
+2\pounds_{\xi_{(1)}}\delta
g_{\mu\nu}+\pounds^2_{\xi_{(1)}}g^{(0)}_{\mu\nu}+
\pounds_{\xi_{(2)}}g^{(0)}_{\mu\nu}\;.
\end{equation}
This leads to the following transformations in the second-order
quantities appearing in Eqs.\ (\ref{eq:m1})--(\ref{eq:m3}):\\

\noindent
{\bf lapse perturbation}
\begin{eqnarray}\label{eq:psi2}
\tilde{\psi}^{(2)} & = & \psi^{(2)}
+\alpha^{(1)}\left[ 2\left(\psi_{(1)}^{\prime} 
+ 2\frac{a'}{a}\psi_{(1)}\right)
+ \alpha_{(1)}^{\prime\prime}  
+5\frac{a'}{a}\alpha_{(1)}^{\prime}
+\left(\frac{a''}{a} 
+\frac{a^{\prime 2}}{a^2}\right) \alpha_{(1)}\right] 
\nonumber \\
& &
+\xi_{(1)}^i\left(2\psi^{(1)}_{,i} + \alpha^{(1)\prime}_{,i}
+\frac{a'}{a}\alpha^{(1)}_{,i}\right)
+2\alpha_{(1)}^{\prime}\left(2\psi_{(1)} 
+\alpha_{(1)}^{\prime}\right) 
\\ & & 
+\xi^{i\prime}_{(1)}\left(\alpha^{(1)}_{,i} -\xi^{(1)\prime}_{i}
-2\omega^{(1)}_i \right)
+\alpha_{(2)}^{\prime}+\frac{a'}{a}\alpha_{(2)}\; ;\nonumber
\end{eqnarray}
{\bf shift perturbation}
\begin{eqnarray}\label{eq:omega2}
\tilde{\omega}^{(2)}_i & = & \omega^{(2)}_i
-4\psi^{(1)}\alpha^{(1)}_{,i}
+\alpha^{(1)}\left[
2\left(\omega^{(1)\prime}_i +2\frac{a'}{a}\omega^{(1)}_i\right)
-\alpha^{(1)\prime}_{,i} + \xi^{(1)\prime\prime}_{i}
-4\frac{a'}{a}\left(\alpha^{(1)}_{,i}
 -\xi^{(1)\prime}_i\right)\right]  
\nonumber \\
& &
+\xi^j_{(1)}\left(2\omega^{(1)}_{i,j}-\alpha^{(1)}_{,ij}
+\xi^{(1)\prime}_{i,j}\right) 
+\alpha_{(1)}^{\prime}\left( 2\omega^{(1)}_i
-3\alpha^{(1)}_{,i}  
+\xi^{(1)\prime}_{i}\right) \\
& & 
+\xi^{j\prime}_{(1)}\left(-4\phi^{(1)}\delta_{ij}+2\chi^{(1)}_{ij}
+2\xi^{(1)}_{j,i}+\xi^{(1)}_{i,j}\right)
+\xi^j_{(1),i}\left(2\omega^{(1)}_j-\alpha^{(1)}_{,j}\right)
-\alpha^{(2)}_{,i} +\xi^{(2)\prime}_{i}\; ;\nonumber
\end{eqnarray}
{\bf spatial metric, trace}
\begin{eqnarray}\label{eq:phi2}
\tilde{\phi}^{(2)} & = & \phi^{(2)}
+\alpha^{(1)}\left[
2\left(\phi_{(1)}^{\prime}+
2\frac{a'}{a}\phi_{(1)}\right)
-\left(\frac{a''}{a}+\frac{a^{\prime 2}}{a^2}\right)\alpha_{(1)}
-\frac{a'}{a} \alpha_{(1)}^{\prime}\right]
+\xi^i_{(1)}\left( 2\phi^{(1)}_{,i}-
\frac{a'}{a}\alpha^{(1)}_{,i}\right)
\nonumber \\
& & 
-\frac{1}{3}\left(-4\phi_{(1)}+\alpha_{(1)}\partial_0
+\xi^i_{(1)}\partial_i
+4\frac{a'}{a}\alpha_{(1)}\right)\nabla^2\beta_{(1)}
-\frac{1}{3}\left(2\omega^i_{(1)}-\alpha^{,i}_{(1)}
+\xi^{i\prime}_{(1)}\right)\alpha^{(1)}_{,i} 
 \\
& & 
-\frac{1}{3}\left(2\chi^{(1)}_{ij}+\xi^{(1)}_{i,j}+
\xi^{(1)}_{j,i}\right)
\xi^{j,i}_{(1)}
-\frac{a'}{a}\alpha_{(2)} -\frac{1}{3}\nabla^2\beta_{(2)}\;
 ;\nonumber
\end{eqnarray}
{\bf spatial metric, traceless part}
\begin{eqnarray}\label{eq:chi2}
\tilde{\chi}^{(2)}_{ij} & = &\chi^{(2)}_{ij}
+2\left(\chi^{(1)\prime}_{ij}
+2\frac{a'}{a}\chi^{(1)}_{ij}\right)\alpha_{(1)}
+2\chi^{(1)}_{ij,k}\xi^k_{(1)} 
\nonumber \\
& &
+2\left(-4\phi_{(1)}+\alpha_{(1)}\partial_0
+\xi^k_{(1)}\partial_k+4\frac{a'}{a}\alpha_{(1)}\right)
\left(d^{(1)}_{(i,j)}+{\rm D}_{ij}\beta_{(1)}\right) 
 \nonumber \\
& &
+2\left[\left(2\omega^{(1)}_{(i}-\alpha^{(1)}_{,(i}
+\xi^{(1)\prime}_{(i}\right) \alpha^{(1)}_{,j)}
-\frac{1}{3}\delta_{ij}
\left(2\omega^k_{(1)}-\alpha^{,k}_{(1)}
+\xi^{k\prime}_{(1)}\right)\alpha^{(1)}_{,k}\right]
 \\
& & 
+2\left[\left(2\chi^{(1)}_{(i|k|}
+\xi^{(1)}_{k,(i}+\xi^{(1)}_{(i,|k|}\right) \xi^{(1)k}_{,j)}
-\frac{1}{3}\delta_{ij}\left(2\chi^{(1)}_{lk}
+\xi^{(1)}_{k,l}+\xi^{(1)}_{l,k}\right) \xi_{(1)}^{k,l}\right]
 \nonumber \\
& &
+2\left(d^{(2)}_{(i,j)} +{\rm D}_{ij}\beta_{(2)}\right)\;.
\nonumber
\end{eqnarray}

For the energy density $\varrho$, or any other scalar,
 we have from Eq.\ (\ref{eq:t2}):
\begin{equation}
\delta^2\tilde{\varrho}=\delta^2\varrho +\left(\pounds_{\xi_{(2)}} +
\pounds^2_{\xi_{(1)}}\right)\varrho_{(0)}
+2\pounds_{\xi_{(1)}}\delta\varrho\;.
\end{equation}
>From this we obtain, using Eq.\ (\ref{eq:liedev1}):
\begin{equation}\label{eq:mut2}
 \delta^2\tilde{\varrho}=\delta^2\varrho +\varrho_{(0)}^\prime\alpha_{(2)} +
\alpha_{(1)}\left(\varrho_{(0)}^{\prime\prime}\alpha_{(1)}
+\varrho_{(0)}^\prime\alpha_{(1)}^\prime+2\delta\varrho^\prime\right)
+\xi^i_{(1)}\left(\varrho^\prime_{(0)}\alpha^{(1)}_{,i}
+2\delta\varrho_{,i}\right)\; .
\end{equation}.

For the 4-velocity $u^\mu$,  we have from
(\ref{eq:t2}): 
\begin{equation}
\delta^2 \tilde{u}^\mu =\delta^2u^\mu+\left(\pounds_{\xi_{(2)}} +
\pounds^2_{\xi_{(1)}}\right)u^\mu_{(0)}
+2\pounds_{\xi_{(1)}}\delta u^\mu\;
.
\end{equation}
Using Eqs.\ (\ref{eq:4v}) and (\ref{eq:liedev2}) this gives:
\begin{eqnarray}\label{eq:v02}
\tilde{v}^0_{(2)} & = &
v^0_{(2)}-\frac{a'}{a}\alpha_{(2)}-\alpha_{(2)}^\prime
+\alpha_{(1)}\left[2\left(v^{0\prime}_{(1)}-
\frac{a'}{a}v^0_{(1)}\right)+
\left(2\frac{a^{\prime
2}}{a^2}-\frac{a^{\prime\prime}}{a}\right)\alpha_{(1)}
+\frac{a'}{a}\alpha^\prime_{(1)}-
\alpha^{\prime\prime}_{(1)}\right]
\nonumber \\
& & +\xi^i_{(1)}\left(2 v^0_{(1),i}
-\frac{a'}{a}\alpha^{(1)}_{,i}-\alpha^{(1)\prime}_{,i}\right)
+\alpha^\prime_{(1)}\left(\alpha^\prime_{(1)}-2v^0_{(1)}\right) 
-2\alpha^{(1)}_{,i}v^i_{(1)}
+\alpha^{(1)}_{,i}\xi^{i\prime}_{(1)}\; ;
\\ \label{eq:vi2}
\tilde{v}^i_{(2)} & = & v^i_{(2)} 
 -\beta_{(2)}^{\prime ,i} - d_{(2)}^{i\prime}
+\alpha_{(1)}\left[
2\left(v^{i\prime}_{(1)}-\frac{a'}{a}v^i_{(1)}\right)
-\left(\xi^{i\prime\prime}_{(1)}
-2\frac{a'}{a}\xi^{i\prime}_{(1)}\right)\right]\nonumber
\\ & & 
+\xi^j_{(1)}\left(2v^i_{(1),j}-\xi^{i\prime}_{(1),j}\right)
-\xi^i_{(1),j} \left( 2 v^j_{(1)} -\xi^{j\prime}_{(1)}\right)
+\xi^{i\prime}_{(1)} 
\left( 2 \psi_{(1)} +\alpha_{(1)}^\prime\right)\;;
\end{eqnarray}
for the time and the space  components respectively.
Again, the 4-velocity $u^\mu$ is subject to $u^\mu u^\nu g_{\mu\nu}=-1$, which
gives Eq.\ (\ref{eq:v0psi2});
 therefore Eq.\ (\ref{eq:v02}) reduces to Eq.\ (\ref{eq:psi2}).

\section{Evolution in the synchronous gauge}

\subsection{General formalism}

In this section we will obtain the second-order perturbations of the
Einstein-de Sitter cosmological model in the synchronous gauge, 
including scalar and tensor modes in the initial conditions. 
The synchronous gauge, that has been one of the most 
frequently used in cosmological perturbation theory,
is defined by the conditions $g_{00}=-a(\tau)^2$ and 
$g_{0i}=0$\cite{bi:li46}. In this way the four degrees of freedom associated
with the coordinate  invariance of the theory are
fixed. 

We start by writing the Einstein's equations for a perfect fluid of 
irrotational dust in synchronous and comoving coordinates. The formalism 
outlined in this subsection is discussed in greater detail in
Ref.\cite{bi:MT}. 
With the purpose of studying gravitational instability in the 
Einstein-de Sitter background, we first factor out the homogeneous and 
isotropic expansion of the universe. 

The line--element is written in the form 
\begin{equation}
ds^2 = a^2(\tau)\big[ - d\tau^2 + \gamma_{ij}({\bf x}, \tau) 
dx^i dx^j \big] \;, 
\end{equation} 
with the spatial coordinates ${\bf x}$ representing Lagrangian coordinates for 
the fluid elements. The scale--factor $a(\tau)\propto \tau^2$ is 
the solution of the Friedmann equations for a perfect fluid of dust 
in the Einstein-de Sitter universe. 

By subtracting the isotropic Hubble--flow, one introduces the extrinsic 
curvature of constant $\tau$ hypersurfaces, 
\begin{equation}
\vartheta^i_{~j} = {1 \over 2} \gamma^{ik}
\gamma_{kj}' \;, 
\end{equation}
with a prime denoting differentiation with respect to the conformal time 
$\tau$. 

One can then write the Einstein's equations in a cosmologically convenient 
form. The energy constraint reads 
\begin{equation}
\vartheta^2 - \vartheta^i_{~j} \vartheta^j_{~i} + {8 \over \tau} 
\vartheta + {\cal R} = {24 \over \tau^2} \delta \;,
\end{equation}
where ${\cal R}^i_{~j}(\gamma)$ is the intrinsic curvature
of constant time hypersurfaces, i.e. 
the conformal Ricci curvature of the three--space with 
metric $\gamma_{ij}$, and ${\cal R}= {\cal R}^i_{~i}$. 
We also introduced the density contrast $\delta \equiv (\varrho - 
\varrho_{(0)}) 
/\varrho_{(0)}$, with $\varrho({\bf x},\tau)$ the mass density and 
$\varrho_{(0)}(\tau) = 3/2\pi G a^2(\tau)\tau^2$ its background 
mean value. 

The momentum constraint reads 
\begin{equation}
\vartheta^i_{~j|i} = \vartheta_{,j} \;,
\end{equation}
where the vertical bar indicates a covariant derivative in the three--space 
with metric $\gamma_{ij}$. 

Finally, after replacing the density from the energy constraint and
subtracting the background contribution, the evolution equation for the 
extrinsic curvature reads 
\begin{equation}
{\vartheta^i_{~j}}' + {4 \over \tau} \vartheta^i_{~j} + 
\vartheta \vartheta^i_{~j} + {1 \over 4} 
\biggl( \vartheta^k_{~\ell} \vartheta^\ell_{~k} - \vartheta^2 \biggr) 
\delta^i_{~j} + {\cal R}^i_{~j} 
- {1 \over 4} {\cal R} \delta^i_{~j} 
= 0 \;. 
\end{equation}

Also useful is the Raychaudhuri equation for the evolution of the 
peculiar volume expansion scalar $\vartheta$, namely
\begin{equation}
\vartheta' + {2 \over \tau } \vartheta + \vartheta^i_{~j} \vartheta^j_{~i} 
+ {6 \over \tau^2} \delta = 0 \;. 
\end{equation}
An advantage of this gauge is that there are only geometric quantities 
in the equations, namely the spatial metric tensor with its time and space 
derivatives. The only remaining variable, the 
density contrast, can indeed be rewritten in terms of $\gamma_{ij}$, 
by solving the continuity equation. We have 
\begin{equation}
\delta({\bf x}, \tau) = (1 + \delta_0({\bf x})) \bigl[\gamma({\bf x}, 
\tau)/ \gamma_0 ({\bf x}) \bigr]^{-1/2} - 1 \;,
\end{equation}
with $\gamma \equiv {\rm det} ~\gamma_{ij}$. We denote by a subscript
$0$ without parenthesis the initial condition of the referred quantity. 

\subsection{First-order perturbations} 

We are now ready to deal with the equations above at the linear level. 
Let us then write the conformal spatial metric tensor in the form 
\begin{equation}
\gamma_{ij} = \delta_{ij} + \gamma^{(1)}_{{\scriptscriptstyle \rm S}ij} \;.
\end{equation}

According to our general definitions we then write
\begin{equation}
\gamma^{(1)}_{{\scriptscriptstyle \rm S}ij} 
= - 2 \phi_{\scriptscriptstyle \rm S}^{(1)} \delta_{ij} + 
{\rm D}_{ij}\chi^{(1)\|}_{\scriptscriptstyle \rm S}
 +\partial_i\chi^{(1)\bot}_{{\scriptscriptstyle \rm S}j} +
\partial_j\chi^{(1)\bot}_{{\scriptscriptstyle \rm S}i} + 
\chi^{(1)\top}_{ij}\;, 
\end{equation}
with
\begin{equation}
\partial^i\chi^{(1)\bot}_{{\scriptscriptstyle \rm S}i} = 
\chi^{(1)\top i}_{~~i} = 
\partial^i \chi^{(1)\top}_{ij} = 0 \;,
\end{equation}
Recall that at first order 
the tensor modes $\chi^{(1)\top}_{ij}$ are gauge-invariant. 

As it is well known, in linear theory, scalar, vector and tensor modes are 
independent. The equation of motion for the tensor modes 
is obtained by linearizing the traceless part of the 
$\vartheta^i_{~j}$ evolution equation. One has
\begin{equation}
{\chi^{(1)\top}_{ij}}'' +  {4 \over \tau} {\chi^{(1)\top}_{ij}}' 
- \nabla^2 \chi^{(1)\top}_{ij} = 0 \;,
\end{equation}
which is the equation for the free propagation of gravitational 
waves in the Einstein-de Sitter universe. The general solution of this 
equation is 
\begin{equation}
\chi^{\top(1)}_{ij}({\bf x},\tau)=\frac{1}{(2\pi)^3} \int d^3{\bf k}
\exp(i{\bf k}\cdot{\bf x}) \chi^{(1)}_\sigma({\bf k},\tau)
\epsilon^{\sigma}_{ij}(\hat{\bf k}),
\end{equation}
where $\epsilon^{\sigma}_{ij}(\hat{\bf k})$ is the polarization tensor,
with $\sigma$ ranging over the polarization components $+,\times$, and
$\chi^{(1)}_\sigma({\bf k},\tau)$ the amplitudes of the two polarization 
states, whose time evolution can be represented as
\begin{equation}
\label{eq:freegw}
\chi^{(1)}_\sigma({\bf k},\tau) = A(k) a_\sigma({\bf k})
\left(\frac{3 j_1(k\tau)}{k\tau}\right),
\end{equation}
with $j_1$ the spherical Bessel function of order one and 
$a_\sigma({\bf k})$ a zero mean random variable with
auto-correlation function
$\langle a_\sigma({\bf k}) a_{\sigma'}({\bf k'})\rangle =(2\pi)^3
k^{-3} \delta^3({\bf k} + {\bf k'}) \delta_{\sigma\sigma'}$.
The spectrum of the gravitational wave background depends on the
processes by which it was generated, and for example in most
inflationary models, $A(k)$ is nearly scale invariant and proportional
to the Hubble constant during inflation.
 
In the irrotational case the linear vector perturbations 
represent gauge modes which can be set to zero: $\chi^{(1)\bot}_i = 0$. 

The two scalar modes are linked together via the momentum constraint, 
leading to the condition 
\begin{equation}
\phi_{\scriptscriptstyle \rm S}^{(1)} + 
{1\over 6} \nabla^2 \chi^{(1)\|}_{\scriptscriptstyle \rm S} = 
\phi^{(1)}_{{\scriptscriptstyle \rm S}0} 
+ {1\over 6} \nabla^2 \chi^{(1)\|}_{{\scriptscriptstyle \rm S}0} \;. 
\end{equation}

The energy constraint gives
\begin{equation}
\nabla^2 \biggl[ {2 \over \tau} {\chi^{(1)\|}_{\scriptscriptstyle \rm S}}' + 
{6 \over \tau^2} \bigl( \chi^{(1)\|}_{\scriptscriptstyle \rm S} - 
\chi^{(1)\|}_{{\scriptscriptstyle \rm S}0} \bigr) 
+ 2 \phi^{(1)}_{{\scriptscriptstyle \rm S}0} + {1 \over 3} \nabla^2 
\chi^{(1)\|}_{{\scriptscriptstyle \rm S}0} \biggr] = 
{12 \over \tau^2} \delta_0 \;, 
\end{equation}
having consistently assumed $\delta_0\ll 1$. 

The evolution equation also gives an equation for the scalar modes, 
\begin{equation}
{\chi_{\scriptscriptstyle \rm S}^{(1)\|}}''+ 
{4 \over \tau} {\chi_{\scriptscriptstyle \rm S}^{(1)\|}}' + {1 \over 3} 
\nabla^2 \chi^{(1)\|}_{\scriptscriptstyle \rm S} = - 2 
\phi_{\scriptscriptstyle \rm S}^{(1)} \;.
\end{equation}
An equation only for the scalar mode $\chi^{(1)\|}_{\scriptscriptstyle \rm S}$ 
can be obtained by 
combining together the evolution equation and the energy constraint,
\begin{equation}
\nabla^2 \biggl[ {\chi_{\scriptscriptstyle \rm S}^{(1)\|}}'' + 
{2 \over \tau} {\chi_{\scriptscriptstyle \rm S}^{(1)\|}}' 
- {6 \over \tau^2} \bigl( \chi^{(1)\|}_{\scriptscriptstyle \rm S} - 
\chi^{(1)\|}_{{\scriptscriptstyle \rm S}0} \bigr) \biggr] = - 
{12 \over \tau^2} \delta_0 \;. 
\end{equation}

On the other hand, by linearizing the solution of the continuity 
equation, we obtain 
\begin{equation}
\delta_{\scriptscriptstyle \rm S}^{(1)} = \delta_0 - {1 \over 2} 
\nabla^2 \bigl(\chi^{(1)\|}_{\scriptscriptstyle \rm S} - 
\chi^{(1)\|}_{{\scriptscriptstyle \rm S}0} \bigr) \;,
\end{equation}
which replaced in the previous expression gives 
\begin{equation}
{\delta_{\scriptscriptstyle \rm S}^{(1)}}'' + 
{2 \over \tau} {\delta_{\scriptscriptstyle \rm S}^{(1)}}' - {6 \over \tau^2}
\delta_{\scriptscriptstyle \rm S}^{(1)} = 0 \;.
\end{equation}
This is the equation for linear density fluctuation 
(see, e.g., Ref.\cite{bi:pjep80}), whose general solution is straightforward 
to obtain. 

The equations above have been obtained in whole generality; 
one could have used 
instead the well-known residual gauge ambiguity of the synchronous 
coordinates (see, e.g., Refs.\cite{bi:MT,bi:russ}) to simplify 
their form. 
For instance, one could fix $\chi^{(1)\|}_0$ so that 
$\nabla^2 \chi^{(1)\|}_{{\scriptscriptstyle \rm S}0} 
= - 2 \delta_0$, and thus the 
$\chi^{(1)\|}_{\scriptscriptstyle \rm S}$ evolution equation takes the same 
form as that for $\delta$. With such a gauge fixing 
one obtains 
\begin{equation}
\chi^{(1)\|}_{\scriptscriptstyle \rm S}({\bf x},\tau) = 
\chi_+({\bf x}) \tau^2 + 
\chi_-({\bf x}) \tau^{-3} \;,
\end{equation} 
where $\chi_\pm$ set the amplitudes of the growing ($+$) and 
decaying ($-$) modes. 
In what follows, we shall restrict ourselves to the growing 
mode. The effect of the decaying mode on second-order perturbations has 
been considered in Ref.\cite{bi:tomita67} and in Ref.\cite{bi:russ} 
and will not be 
studied here. 
The amplitude of the growing mode is related to the initial {\em peculiar 
gravitational potential}, through $\chi_+ \equiv - {1 \over 3} 
\varphi$, where in turn, $\varphi$ is related to $\delta_0$ through the 
cosmological Poisson equation $\nabla^2 \varphi({\bf x}) = {6 \over \tau_0^2} 
\delta_0({\bf x})$. Therefore, 
\begin{equation}\label{eq:chis}
{\rm D}_{ij}\chi^{(1)\|}_{\scriptscriptstyle \rm S} 
= - {\tau^2 \over 3} \biggl( \varphi_{,ij} 
- {1 \over 3} \delta_{ij} \nabla^2 \varphi \biggr) \;.
\end{equation}
The remaining scalar mode 
\begin{equation}
\phi_{\scriptscriptstyle \rm S}^{(1)}({\bf x},\tau) = 
{5 \over 3} \varphi({\bf x}) + {\tau^2 \over 18} 
\nabla^2 \varphi({\bf x}) \;
\end{equation}
immediately follows. 

The linear metric perturbation therefore reads 
\begin{equation}
\gamma^{(1)}_{{\scriptscriptstyle \rm S}ij} = 
- {10 \over 3} \varphi \delta_{ij} - {\tau^2  \over 3} 
\varphi_{,ij} + \chi^{(1)\top}_{ij} \;.
\end{equation} 

With purely growing-mode initial conditions, the linear density contrast reads 
\begin{equation}
\delta_{\scriptscriptstyle \rm S}^{(1)} = {\tau^2 \over 6} \nabla^2 \varphi \;. 
\end{equation}

\subsection{Second-order perturbations} 

The conformal spatial metric tensor up to second order is expanded as 
\begin{equation} 
\gamma_{ij} = \delta_{ij} + \gamma^{(1)}_{{\scriptscriptstyle \rm S}ij} 
+ {1 \over 2} 
\gamma^{(2)}_{{\scriptscriptstyle \rm S}ij} \;,
\end{equation}
with 
\begin{equation} 
\gamma^{(2)}_{{\scriptscriptstyle \rm S}ij} = - 2 
\phi_{\scriptscriptstyle \rm S}^{(2)} 
\delta_{ij} + \chi^{(2)}_{{\scriptscriptstyle \rm S}ij} 
\end{equation}
and $\chi^{(2)i}_{{\scriptscriptstyle \rm S}~~~j}=0$. 

The technique of second-order perturbation theory is straightforward: with the 
help of the relations reported in Appendix B, we first 
substitute the expansion above in our exact fluid-dynamical equations 
(momentum and energy constraints plus evolution and Raychaudhuri 
equations) obtaining equations for 
$\gamma^{(2)}_{{\scriptscriptstyle \rm S}ij}$ with source terms 
containing quadratic combinations of 
$\gamma^{(1)}_{{\scriptscriptstyle \rm S}ij}$ plus a few more terms 
involving $\delta_0$. Next, we have to solve these equations for the 
modes $\phi_{\scriptscriptstyle \rm S}^{(2)}$ and 
$\chi^{(2)}_{{\scriptscriptstyle \rm S}ij}$ 
in terms of the initial 
peculiar gravitational potential $\varphi$ and the linear tensor modes 
$\chi^{(1)\top}_{ij}$. 

Let us now give the equations which govern the evolution of the second-order 
metric perturbations. 
\\

\noindent
{\bf Raychaudhuri equation} 

\begin{eqnarray} 
{\phi_{\scriptscriptstyle \rm S}^{(2)}}'' && + 
{2 \over \tau} {\phi_{\scriptscriptstyle \rm S}^{(2)}}' - {6 \over \tau^2} 
\phi_{\scriptscriptstyle \rm S}^{(2)} = - {1 \over 6} {\gamma^{(1)ij}_S}' 
\biggl( {\gamma^{(1)}_{{\scriptscriptstyle \rm S}ij}}' 
- { 4 \over \tau} \gamma^{(1)}_{{\scriptscriptstyle \rm S}ij} \biggr) 
+ {1 \over 6} \biggl[ 2 g^{(1)ij}_{\scriptscriptstyle \rm S} 
\biggl( 2 \gamma^{(1)k}_{{\scriptscriptstyle \rm S}i~,kj} 
- \nabla^2 \gamma^{(1)}_{{\scriptscriptstyle \rm S}ij} - 
\gamma^{(1)k}_{{\scriptscriptstyle \rm S}k~,ij} \biggr)
\nonumber\\
&& - \gamma^{(1)k}_{{\scriptscriptstyle \rm S}k} 
\biggl(\gamma^{(1)ij}_{{\scriptscriptstyle \rm S}~~~,ij} - \nabla^2 
\gamma^{(1)i}_{{\scriptscriptstyle \rm S}i} \biggr) 
\biggr] - {2 \over \tau^2} \biggl[ - {1 \over 4} 
\biggl(\gamma^{(1)i}_{{\scriptscriptstyle \rm S}i} 
- \gamma^{(1)~i}_{{\scriptscriptstyle \rm S}0i} \biggr)^2
\nonumber\\
&&
- {1 \over 2} 
\biggl(\gamma^{(1)ij}_{\scriptscriptstyle \rm S} 
\gamma^{(1)}_{{\scriptscriptstyle \rm S}ij} - 
\gamma^{(1)ij}_{{\scriptscriptstyle \rm S}0} 
\gamma^{(1)}_{{\scriptscriptstyle \rm S}0ij} \biggr) 
+ \delta_0 \biggl(\gamma^{(1)i}_{{\scriptscriptstyle \rm S}i} - 
\gamma^{(1)~i}_{{\scriptscriptstyle \rm S}0i} \biggr) \biggr] 
\;; 
\end{eqnarray} 

\noindent
{\bf energy constraint} 

\begin{eqnarray}
{2 \over \tau} {\phi^{(2)}_{\scriptscriptstyle \rm S}}' && - {1 \over 3} 
\nabla^2 \phi^{(2)}_{\scriptscriptstyle \rm S} +{6 \over \tau^2} 
\phi^{(2)}_{\scriptscriptstyle \rm S} - {1 \over 12} 
\chi^{(2)ij}_{{\scriptscriptstyle \rm S}~~~,ij} 
= - {2 \over 3 \tau} \gamma^{(1)ij}_{\scriptscriptstyle \rm S} 
{\gamma^{(1)}_{{\scriptscriptstyle \rm S}ij}}' - 
{1 \over 24} \biggl( {\gamma^{(1)ij}_{\scriptscriptstyle \rm S}}' 
{\gamma^{(1)}_{{\scriptscriptstyle \rm S}ij}}' - 
{\gamma^{(1)i}_{{\scriptscriptstyle \rm S}i}}'
{\gamma^{(1)j}_{{\scriptscriptstyle \rm S}j}}'\biggr) 
\nonumber\\ 
&& 
+ 
{1 \over 6} \biggl[ \gamma^{(1)ij}_{\scriptscriptstyle \rm S}
 \biggl( \nabla^2 \gamma^{(1)}_{{\scriptscriptstyle \rm S}ij}
+ \gamma^{(1)k}_{{\scriptscriptstyle \rm S}k~,ij} - 
2 \gamma^{(1)k}_{{\scriptscriptstyle \rm S}i~,jk} \biggr) + 
\gamma^{(1)ki}_{{\scriptscriptstyle \rm S}~~~,k} 
\biggl( \gamma^{(1)j}_{{\scriptscriptstyle \rm S}j~,i} - 
\gamma^{(1)j}_{{\scriptscriptstyle \rm S}i~,j} \biggr) 
\nonumber\\ 
&& 
+ {3 \over 4} \gamma^{(1)ij}_{{\scriptscriptstyle \rm S}~~~,k} 
\gamma^{(1),k}_{{\scriptscriptstyle \rm S}ij} 
- {1 \over 2} \gamma^{(1)ij}_{{\scriptscriptstyle \rm S}~~~,k} 
\gamma^{(1)k}_{{\scriptscriptstyle \rm S}i~~,j}  
- {1 \over 4} \gamma^{(1)i,k}_{{\scriptscriptstyle \rm S}i} 
\gamma^{(1)j}_{{\scriptscriptstyle \rm S}j~,k} \biggr] 
\nonumber\\ 
&& + {2 \over \tau^2} \biggl[ - 
{1 \over 4} \biggl(\gamma^{(1)i}_{{\scriptscriptstyle \rm S}i} 
- \gamma^{(1)~i}_{{\scriptscriptstyle \rm S}0i} \biggr)^2 - {1 \over 2} 
\biggl( \gamma^{(1)ij}_{\scriptscriptstyle \rm S} 
\gamma^{(1)}_{{\scriptscriptstyle \rm S}ij} - 
\gamma^{(1)ij}_{{\scriptscriptstyle \rm S}0} 
\gamma^{(1)}_{{\scriptscriptstyle \rm S}0ij} \biggr) 
+ \delta_0 \biggl(\gamma^{(1)i}_{{\scriptscriptstyle \rm S}i} - 
\gamma^{(1)~i}_{{\scriptscriptstyle \rm S}0i} \biggr) \biggr] 
\;; 
\end{eqnarray} 

\noindent
{\bf momentum constraint} 

\begin{equation}
2 {\phi^{(2)}_{{\scriptscriptstyle \rm S},j}}' + {1 \over 2} 
{\chi^{(2)i}_{{\scriptscriptstyle \rm S}j~,i}}' = 
\gamma^{(1)ik}_{\scriptscriptstyle \rm S} 
\biggl( {\gamma^{(1)}_{{\scriptscriptstyle \rm S}jk,i}}' - 
{\gamma^{(1)}_{{\scriptscriptstyle \rm S}ik,j}}' \biggr) 
+ \gamma^{(1)ik}_{{\scriptscriptstyle \rm S}~~~~,i} 
{\gamma^{(1)}_{{\scriptscriptstyle \rm S}jk}}' 
- {1 \over 2} \gamma^{(1)ik}_{{\scriptscriptstyle \rm S}~~~~,j} 
{\gamma^{(1)}_{{\scriptscriptstyle \rm S}ik}}' 
- {1 \over 2} \gamma^{(1)i}_{{\scriptscriptstyle \rm S}i~,k} 
{\gamma^{(1)k}_{{\scriptscriptstyle \rm S}j}}' 
\;; 
\end{equation} 

\noindent
{\bf evolution equation} 

\begin{eqnarray}
- \biggl({\phi^{(2)}_{\scriptscriptstyle \rm S}}'' && + {4 \over \tau} 
{\phi^{(2)}_{\scriptscriptstyle \rm S}}' \biggr) \delta^i_{~j} + 
{1 \over 2} \biggl( {\chi^{(2)i}_{{\scriptscriptstyle \rm S}j}}'' + 
{4 \over \tau} {\chi^{(2)i}_{{\scriptscriptstyle \rm S}j}}' \biggr) + 
\phi^{(2),i}_{{\scriptscriptstyle \rm S},j} - {1 \over 4} 
\chi^{(2)k\ell}_{{\scriptscriptstyle \rm S}~~~,k\ell} \delta^i_{~j} + 
{1 \over 2} \chi^{(2)ki}_{{\scriptscriptstyle \rm S}~~~,kj} + 
{1 \over 2} \chi^{(2)k,i}_{{\scriptscriptstyle \rm S}j~~~,k} - {1 \over 2} 
\nabla^2 \chi^{(2)i}_{{\scriptscriptstyle \rm S}j} 
\nonumber\\ 
&& = {\gamma^{(1)ik}_{\scriptscriptstyle \rm S}}' 
{\gamma^{(1)}_{{\scriptscriptstyle \rm S}kj}}' - {1 \over 2} 
{\gamma^{(1)k}_{{\scriptscriptstyle \rm S}k}}' 
{\gamma^{(1)i}_{{\scriptscriptstyle \rm S}j}}' + {1 \over 8} 
\biggl[\biggl({\gamma^{(1)k}_{{\scriptscriptstyle \rm S}k}}'\biggr)^2 - 
{\gamma^{(1)k}_{{\scriptscriptstyle \rm S}\ell}}' 
{\gamma^{(1)\ell}_{{\scriptscriptstyle \rm S}k}}' \biggr] \delta^i_{~j} 
- {1 \over 2} \biggl[ - \gamma^{(1)i}_{{\scriptscriptstyle \rm S}j} 
\biggl( \gamma^{(1)k,\ell}_{{\scriptscriptstyle \rm S}\ell~~~,k}
- \nabla^2 \gamma^{(1)k}_{{\scriptscriptstyle \rm S}k} \biggr) 
\nonumber\\ 
&& + 2 \gamma^{(1)k\ell}_{\scriptscriptstyle \rm S} 
\biggl( \gamma^{(1)i}_{{\scriptscriptstyle \rm S}j~,k\ell} + 
\gamma^{(1)~,i}_{{\scriptscriptstyle \rm S}k\ell~,j} 
- \gamma^{(1)i}_{{\scriptscriptstyle \rm S}\ell~,jk}  
- \gamma^{(1)~,i}_{{\scriptscriptstyle \rm S}\ell j~,k}  \biggr) 
+ 2 \gamma^{(1)k\ell}_{{\scriptscriptstyle \rm S}~~~,k} 
\biggl( \gamma^{(1)i}_{{\scriptscriptstyle \rm S}j~,\ell} 
- \gamma^{(1)i}_{{\scriptscriptstyle \rm S}\ell~,j} - 
\gamma^{(1),i}_{{\scriptscriptstyle \rm S}j\ell} \biggr) 
\nonumber\\ 
&& 
+ 2 \gamma^{(1)ki}_{{\scriptscriptstyle \rm S}~~~,\ell} 
\gamma^{(1),\ell}_{{\scriptscriptstyle \rm S}jk} 
- 2 \gamma^{(1)ki}_{{\scriptscriptstyle \rm S}~~~,\ell} 
\gamma^{(1)\ell}_{{\scriptscriptstyle \rm S}j~,k} 
+ \gamma^{(1)k\ell}_{{\scriptscriptstyle \rm S}~~~,j} 
\gamma^{(1)~,i}_{{\scriptscriptstyle \rm S}k\ell} 
+ \gamma^{(1)\ell}_{{\scriptscriptstyle \rm S}\ell~,k} 
\biggl( \gamma^{(1)ki}_{{\scriptscriptstyle \rm S}~~~,j} + 
\gamma^{(1)k,i}_{{\scriptscriptstyle \rm S}j} - 
\gamma^{(1)i,k}_{{\scriptscriptstyle \rm S}j} \biggr) 
\nonumber\\ 
&& 
- \gamma^{(1)k\ell}_{\scriptscriptstyle \rm S} 
\biggl( \nabla^2 \gamma^{(1)}_{{\scriptscriptstyle \rm S}k\ell} 
+ \gamma^{(1)m}_{{\scriptscriptstyle \rm S}m~,k\ell} - 
2 \gamma^{(1)m}_{{\scriptscriptstyle \rm S}k~~,m\ell} \biggr) 
\delta^i_{~j}  
- \gamma^{(1)\ell k}_{{\scriptscriptstyle \rm S}~~~,\ell} 
\biggl( \gamma^{(1)m}_{{\scriptscriptstyle \rm S}m~,k} - 
\gamma^{(1)m}_{{\scriptscriptstyle \rm S}k~,m} 
\biggr) \delta^i_{~j} 
\nonumber\\ 
&& 
- {3 \over 4} 
\gamma^{(1)k\ell}_{{\scriptscriptstyle \rm S}~~~,m} 
\gamma^{(1),m }_{{\scriptscriptstyle \rm S}k\ell} 
\delta^i_{~j} 
+ {1 \over 2} \gamma^{(1)k\ell}_{{\scriptscriptstyle \rm S}~~~,m} 
\gamma^{(1)m }_{{\scriptscriptstyle \rm S}k~~,\ell} 
\delta^i_{~j} + {1 \over 4} \gamma^{(1)k,m}_{{\scriptscriptstyle \rm S}k} 
\gamma^{(1)\ell}_{{\scriptscriptstyle \rm S}\ell~,m} 
\delta^i_{~j} 
\;. 
\end{eqnarray} 

The next step is to solve these equations. 
In these calculations, we can make the simplifying assumption that the initial 
conditions are taken at conformal time $\tau_0=0$ (implying also $\delta_0=0$). 
One can start from the 
Raychaudhuri equation, to obtain the trace of the second-order metric tensor. 
(Actually, in order to obtain the sub-leading mode generated by linear 
scalar modes, we also need the energy constraint). The resulting expression 
for $\phi_{\scriptscriptstyle \rm S}^{(2)}$ is 
\begin{equation}
\phi_{\scriptscriptstyle \rm S}^{(2)} = 
\frac{\tau^4}{252}\left(-\frac{10}{3}\varphi^{,ki}
\varphi_{,ki}+(\nabla^2\varphi)^2\right)
+{5 \tau^2 \over 18}\left(\varphi^{,k}\varphi_{,k}+
\frac{4}{3}\varphi\nabla^2\varphi\right) + 
\phi^{(2)}_{{\scriptscriptstyle \rm S}({\rm t})}  \;,
\end{equation}
where $\phi_{{\scriptscriptstyle \rm S}({\rm t})}^{(2)}$, which is the part of 
$\phi_{\scriptscriptstyle \rm S}^{(2)}$ 
generated by the presence of tensor modes at the linear level, 
reads 
\begin{equation} \label{eq:qtau}
\phi_{{\scriptscriptstyle \rm S}({\rm t})}^{(2)} = {\tau^2 \over 5} \int_0^\tau
{d \tau' \over \tau'} {\cal Q}(\tau') - 
{1 \over 5 \tau^3} \int_0^\tau d\tau' \tau'^4 
{\cal Q}(\tau') \;, 
\end{equation}
with ${\cal Q}({\bf x},\tau)$ a source term whose explicit form is reported 
in Appendix C. 

The expression for $\chi^{(2)}_{{\scriptscriptstyle \rm S}ij}$ is 
obtained by first replacing 
$\phi_{\scriptscriptstyle \rm S}^{(2)}$ into the remaining equations and 
solving them in the following 
order: energy constraint $\longrightarrow$ momentum constraint 
$\longrightarrow$ (traceless part of the) evolution equation. 
We obtain 
\begin{eqnarray}
\chi_{{\scriptscriptstyle \rm S}ij}^{(2)}& = 
&\frac{\tau^4}{126}\left(19\varphi^{,k}_{,i}
\varphi_{,kj}-12 \varphi_{,ij} \nabla^2\varphi
+4 (\nabla^2\varphi)^2 \delta_{ij}
-\frac{19}{3}\varphi^{,kl}\varphi_{,kl} \delta_{ij}\right)\nonumber\\
&+&\frac{5\tau^2}{9}\left(-6\varphi_{,i}\varphi_{,j}
-4\varphi \varphi_{,ij}+2 \varphi^{,k}\varphi_{,k}\delta_{ij}
+\frac{4}{3}\varphi\nabla^2\varphi\delta_{ij}\right)
+ \pi_{{\scriptscriptstyle \rm S}ij} + 
\chi^{(2)}_{{\scriptscriptstyle \rm S}({\rm t})ij} \;,
\end{eqnarray}
where $\chi^{(2)}_{{\scriptscriptstyle \rm S}({\rm t})ij}$ is the part of 
the traceless tensor $\chi^{(2)}_{{\scriptscriptstyle \rm S}ij}$ generated 
by the presence of tensor 
modes at the linear level and includes the effects
of scalar-tensor and tensor-tensor couplings; its expression 
can be derived from the equations given in Appendix C. 
The transverse and traceless contribution 
$\pi_{{\scriptscriptstyle \rm S}ij}$, which 
represents the second-order tensor mode generated by scalar initial 
perturbations, is determined by the inhomogeneous wave-equation 
\begin{equation}
\pi_{{\scriptscriptstyle \rm S}ij}'' + 
\frac{4}{\tau}\pi_{{\scriptscriptstyle \rm S}ij}'-
\nabla^2 \pi_{{\scriptscriptstyle \rm S}ij}=
-\frac{\tau^4}{21}\nabla^2 {\cal S}_{ij},
\end{equation}
with
\begin{equation}
{\cal S}_{ij}=\nabla^2 \Psi_0 \delta_{ij}+ \Psi_{0,ij}+
2\left(\varphi_{,ij}\nabla^2\varphi-\varphi_{,ik}
\varphi^{,k}_{,j}\right),
\end{equation}
where
\begin{equation}
\nabla^2 \Psi_0=-\frac{1}{2}\left((\nabla^2\varphi)^2-
\varphi_{,ik}\varphi^{,ik}\right).
\end{equation}
This equation can be solved using the Green method; we obtain for 
$\pi_{{\scriptscriptstyle \rm S}ij}$ that
\begin{equation}
\pi_{ij}({\bf x},\tau) = \frac{\tau^4}{21}{\cal S}_{ij}({\bf x}) + 
\frac{4\tau^2}{3} {\cal T}_{ij}({\bf x})
+\tilde{\pi}_{ij}({\bf x},\tau),
\end{equation}
where $\nabla^2 {\cal T}_{ij} = {\cal S}_{ij}$ and the remaining piece
$\tilde{\pi}_{ij}$, containing a term that is
constant in time and another one that oscillates with decreasing
amplitude, can be written as
\begin{equation}
\label{eq:tildepi}
\tilde{\pi}_{ij}({\bf x},\tau)=\frac{1}{(2\pi)^3} \int d^3{\bf k}
\exp(i{\bf k}\cdot{\bf x})\frac{40}{k^4} {\cal S}_{ij}({\bf k})\left(
\frac{1}{3} - {j_1(k\tau) \over k\tau} \right),
\end{equation}
with ${\cal S}_{ij}({\bf k})=\int d^3{\bf x}\exp(-i{\bf k}\cdot{\bf x})
{\cal S}_{ij}({\bf x})$.

The second-order density contrast reads 
\begin{eqnarray} 
\delta^{(2)}_{\scriptscriptstyle \rm S} & = & {\tau^4\over 252} 
\biggl(5\bigl(\nabla^2 \varphi\bigr)^2 
+ 2 \varphi^{,ij}\varphi_{,ij} \biggr) + 
{\tau^2 \over 36} \biggl( 15 \varphi^{,i} \varphi_{,i} + 40 
\varphi\nabla^2 \varphi - 6 \varphi^{,ij} \chi_{ij}^{(1)\top} \biggr) \\
\nonumber
&& 
+ {1 \over 4} 
\biggl( \chi^{(1)\top ij} \chi^{(1)\top}_{ij} - 
\chi^{(1)\top ij}_0 \chi^{(1)\top}_{0ij} \biggr) + {3 \over 2} 
\phi^{(2)}_{{\scriptscriptstyle \rm S}({\rm t})}  \;. 
\end{eqnarray}

An important aspect of our results is that linear tensor modes 
(gravitational waves) can generate second-order perturbations of any kind 
(scalars, vectors and tensors). This interesting fact, which has been first 
noticed by Tomita\cite{bi:tomita71-72}, 
is nicely displayed by the above formula for the 
mass-density contrast, which even in the absence of initial density 
fluctuations, takes a contribution from primordial gravitational waves. 
More in general, we should stress that our expressions completely determine 
the rate of growth of perturbations up to second order. 

\setcounter{equation}{0}


\section{From the Synchronous to the Poisson gauge}

In this section we are going to obtain the metric perturbations 
in the Poisson gauge by transforming the results obtained in the
synchronous gauge in the previous section.
 The Poisson gauge, recently discussed by 
Bertschinger\cite{bi:bertschinger} and Ma and Bertschinger\cite{bi:mabe}, 
is defined by ${\omega_i}^{(r),i}={\chi_{ij}}^{(r),j}=0$.  Then, one 
scalar degree
of freedom is eliminated from $g_{0i}$ ($\omega^{(r)\|}=0$), and one
scalar and two vector degrees of freedom from $g_{ij}$
($\chi^{(r)\|}=\chi_i^{(r)\bot}=0$).  This gauge generalizes the
well-known {\em longitudinal gauge\/} to include vector and tensor
modes.  The latter gauge, in which $\omega_i^{(r)}=\chi_{ij}^{(r)}=0$, has
been widely used in the literature to investigate the evolution of
scalar perturbations\cite{bi:MFB}. Since the vector and tensor modes
are set to zero by hand, the longitudinal gauge cannot be used to
study perturbations beyond the linear regime, because in the nonlinear
case the scalar, vector, and tensor modes are dynamically
coupled. In other words, even if one starts with purely
scalar linear perturbations as initial conditions for the second-order
theory, vector and tensor modes are dynamically
generated\cite{bi:MPS}.

\subsection{First-order transformations}

Given the perturbation of the metric in one gauge, it is easy to
obtain, from Eqs.\ (\ref{eq:pertpsi})--(\ref{eq:pertchi}), the gauge
transformation to the other one, hence the perturbations in the new
gauge.  In the particular case of the synchronous and Poisson gauges,
we have:
\begin{equation}
\psi_{\scriptscriptstyle \rm P}^{(1)}
=\alpha^{(1)\prime}+{a'\over a}\,\alpha^{(1)}\;;
\label{eq:i}
\end{equation}
\begin{equation}
\alpha^{(1)}=\beta^{(1)\prime}\;;
\label{eq:iii}
\end{equation}
\begin{equation}
\omega_{{\scriptscriptstyle \rm P}\ i}^{(1)}=d^{(1)\prime}_i\;;
\label{eq:iv}
\end{equation}
\begin{equation}
\phi_{\scriptscriptstyle \rm P}^{(1)}=\phi_{\scriptscriptstyle
\rm S}^{(1)}-{1\over 3}\,\nabla^2\beta^{(1)}-
{a'\over a}\,\alpha^{(1)}\;;
\label{eq:ii}
\end{equation}
\begin{equation}
{\rm D}_{ij}\left(\chi_{\scriptscriptstyle
\rm S}^{(1)\|}+2\beta^{(1)}\right)=0\;;
\label{eq:v}
\end{equation}
\begin{equation}
\chi_{{\scriptscriptstyle
\rm S}\ (i,j)}^{(1)\bot}+d^{(1)}_{(i,j)}=0\;;
\label{vi}
\end{equation}
\begin{equation}
\chi_{{\scriptscriptstyle\rm  P}\ ij}^{(1)\top}=
\chi_{{\scriptscriptstyle
\rm S}\ ij}^{(1)\top}\;.
\label{eq:vii}
\end{equation}

The parameters $\beta^{(1)}$, $\alpha^{(1)}$, and $d^{(1)}_i$ of the
gauge transformation can be obtained from Eqs.\ (\ref{eq:v}), (\ref{eq:iii}),
and (\ref{vi}) respectively, while the transformed metric
perturbations follow from Eqs.\ (\ref{eq:i}), (\ref{eq:iv}), (\ref{eq:ii}), and
(\ref{eq:vii}).

Once these parameters are known, the transformation rules for the
energy density $\varrho$ or any other scalar, and those for the 4-velocity
$u^\mu$, follow trivially from Eqs.\ (\ref{eq:mut}), (\ref{eq:v0}),
and (\ref{eq:vi}). In the irrotational case studied in the last section
$\chi_{{\scriptscriptstyle \rm S}\ ij}^{(1)\bot}=v^i_{(1)\perp}=0$ 
and thus $d_i^{(1)}=\omega^{(1)}_{{\scriptscriptstyle \rm P}\ i}
=\chi_{{\scriptscriptstyle \rm P}\ ij}^{(1)\bot}=0$.

\subsection{Second-order transformations}

 The more general transformation
expressions follow straightforwardly from Eqs.\
(\ref{eq:psi2})--(\ref{eq:chi2}), (\ref{eq:mut2}), and (\ref{eq:vi2}).

Transforming from the synchronous to the Poisson gauge, the expression
for $\psi^{(2)}_{\scriptscriptstyle \rm P}$ can be easily obtained
from Eq.\ (\ref{eq:psi2}), using Eq.\ (\ref{eq:iii}) and the condition
$d_i^{(1)}=0$ to express all the first-order quantities in terms of
$\beta^{(1)}$:
\begin{equation}
\label{eq:psiP}
\psi^{(2)}_{\scriptscriptstyle \rm P}=\beta_{(1)}^{\prime}\left[
 \beta_{(1)}^{\prime\prime\prime}
+5\frac{a'}{a}\beta_{(1)}^{\prime\prime}
+\left(\frac{a''}{a} +\frac{a^{\prime 2}}{a^2}\right) 
\beta_{(1)}^{\prime}\right]
+\beta_{(1)}^{ ,i}\left(
\beta^{(1)\prime\prime}_{,i}
+\frac{a'}{a}\beta^{(1)\prime}_{,i}\right)
+2\beta_{(1)}^{\prime\prime 2} 
+\alpha^{(2)\prime}+\frac{a'}{a}\alpha^{(2)}\;.
\end{equation}

For $\omega^{(2)}_{{\scriptscriptstyle \rm P}\ i}$ and
$\phi^{(2)}_{\scriptscriptstyle \rm P}$ we get:
\begin{equation}\label{eq:omegaP}
\omega^{(2)}_{{\scriptscriptstyle \rm P}\ i}=
-2\left(
2\phi^{(1)}_{\scriptscriptstyle \rm S} 
+\beta_{(1)}^{\prime\prime} 
-\frac{2}{3}\nabla^2\beta_{(1)}\right)\beta^{(1)\prime}_{,i}
-2\beta^{(1)\prime}_{,j}\beta^{(1),j}_{,i}
+2 \chi_{ij}^{(1)\top}\beta_{(1)}^{\prime ,i }
-\alpha^{(2)}_{,i} +\beta^{(2)\prime}_{,i} +d^{(2) \prime}_i\;;
\end{equation}
\begin{eqnarray}\label{eq:phiP}
\phi^{(2)}_{\scriptscriptstyle \rm P}& =
 & \phi^{(2)}_{\scriptscriptstyle
\rm S}
+\beta^{\prime}_{(1)}\left[
2\left(\phi_{\scriptscriptstyle
\rm S}^{(1)\prime}+
2\frac{a'}{a}\phi^{(1)}_{\scriptscriptstyle
\rm S}\right)
-\left(\frac{a''}{a}
+\frac{a^{\prime 2}}{a^2}\right)\beta^{\prime}_{(1)}
-\frac{a'}{a} \beta_{(1)}^{\prime\prime}\right]
\nonumber \\
& &
-\frac{1}{3}\left(-4\phi^{(1)}_{\scriptscriptstyle
\rm S}+\beta_{(1)}^{\prime}\partial_0
+\beta^{,i}_{(1)}\partial_i
+4\frac{a'}{a}\beta^{\prime}_{(1)}
+\frac{4}{3}\nabla^2\beta_{(1)}\right)\nabla^2\beta_{(1)}
\\ 
&& \nonumber
+\beta^{,i}_{(1)}\left( 2\phi^{(1)}_{{\scriptscriptstyle
\rm S} ,i}-\frac{a'}{a}\beta^{(1)\prime}_{,i}\right)
+\frac{2}{3}\beta^{(1)}_{,ij}\beta^{,ij}_{(1)}
-\frac{2}{3}\chi_{ij}^{(1)\top}\beta_{(1)}^{,ij}
-\frac{a'}{a}\alpha_{(2)} -\frac{1}{3}\nabla^2\beta_{(2)}\;.
\end{eqnarray}

For $\chi^{(2)}_{{\scriptscriptstyle \rm P}\ ij}$ we obtain:
\begin{eqnarray}
\chi^{(2)}_{{\scriptscriptstyle \rm P}\ ij} & = & 
\chi^{(2)}_{{\scriptscriptstyle \rm S}\ ij}
+2\left(\frac{4}{3}\nabla^2\beta_{(1)}
-4\phi^{(1)}_{\scriptscriptstyle
\rm S} -
\beta^{\prime}_{(1)}\partial_0-\beta_{(1)}^{,k}\partial_k\right){\rm
D}_{ij} \beta_{(1)}
 \nonumber \\
& & \label{eq:chiP}
-4\left(\beta^{(1)}_{,ik}\beta^{,k}_{(1),j}
-\frac{1}{3}\delta_{ij}\beta^{(1)}_{,lk}\beta_{(1)}^{,lk}\right)
+2\left(\chi_{ij}^{(1)\top \prime}+2\frac{a'}{a}\chi_{ij}^{(1)\top}
\right)\beta^{(1)\prime}
   \\
& & +2\chi_{ij,k}^{(1)\top}\beta^{(1),k}
+2\chi_{ik}^{(1)\top}\beta^{(1),k}_{,j}
+2\chi_{jk}^{(1)\top}\beta^{(1),k}_{,i}
  \nonumber \\
& &-\frac{4}{3}\delta_{ij}\chi_{lk}^{(1)\top}\beta^{(1),lk}
+2\left(d^{(2)}_{(i,j)} +{\rm D}_{ij}\beta^{(2)}\right)\; .
\nonumber
\end{eqnarray} 

Given the metric perturbations in the synchronous gauge, these
constitute a set of coupled equations for the second-order parameters
of the transformation, $\alpha^{(2)},\beta^{(2)}$, and $d_i^{(2)}$,
and the second-order metric perturbations in the Poisson gauge,
$\psi_{\scriptscriptstyle \rm P}^{(2)}$, $\omega_{{\scriptscriptstyle
\rm P}\ i}^{(2)}$, $\phi_{\scriptscriptstyle \rm P}^{(2)}$, and
$\chi^{(2)}_{{\scriptscriptstyle \rm P}\ ij}$.  This system can be
solved in the following way.  Since in the Poisson gauge
$\partial^i\chi^{(2)}_{{\scriptscriptstyle \rm P}\ ij}=0$, we can use
the fact that $\partial^i \partial^j \chi^{(2)}_{{\scriptscriptstyle
\rm P}\ ij}=0$ and the property $\partial^id_i^{(1)}=0$, together with
Eq.\ (\ref{eq:chiP}), to obtain an expression for
$\nabla^2\nabla^2\beta^{(2)}$, from which $\beta^{(2)}$ can be
computed:
\begin{eqnarray}\label{eq:beta2}
\nabla^2\nabla^2\beta^{(2)} & = & -\frac{3}{4}
\chi^{(2),ij}_{{\scriptscriptstyle \rm S}\ ij}+
6\phi_{\scriptscriptstyle \rm S}^{(1),ij}
\beta^{(1)}_{,ij} -2 \nabla^2\phi_{\scriptscriptstyle \rm S}^{(1)} 
\nabla^2\beta_{(1)}
+8  \phi_{\scriptscriptstyle \rm S}^{(1),i} \nabla^2\beta^{(1)}_{,i}
+4 \phi_{\scriptscriptstyle \rm S}^{(1)} \nabla^2\nabla^2\beta_{(1)} 
 \nonumber \\
&&
+4\nabla^2\beta^{(1)}_{,ij}\beta_{(1)}^{,ij}
-\frac{1}{6}\nabla^2\beta_{(1)}^{,i} \nabla^2\beta^{(1)}_{,i}
+\frac{5}{2}\beta_{(1)}^{,ijk} \beta^{(1)}_{,ijk}
-\frac{2}{3}\nabla^2\beta_{(1)} \nabla^2\nabla^2\beta_{(1)}
\\ 
&&
+\frac{3}{2}\beta_{(1)}^{,ij\prime} \beta^{(1)\prime}_{,ij} 
-\frac{1}{2}\nabla^2\beta_{(1)}^{\prime} \nabla^2\beta_{(1)}^{\prime}
+2\beta_{(1)}^{,i\prime} \nabla^2\beta^{(1)\prime}_{,i} 
+\beta_{(1)}^{\prime} \nabla^2\nabla^2\beta_{(1)}^{\prime} 
+\beta_{(1)}^{,i} \nabla^2\nabla^2\beta^{(1)}_{,i}
 \nonumber \\
&&-\frac{3}{2}\left(\chi_{ij}^{(1)\top \prime}+2\frac{a'}{a}
\chi_{ij}^{(1)\top}\right)\beta^{(1)\prime,ij}
-\frac{5}{2}\chi_{ij,k}^{(1)\top} \beta^{(1),ijk}
-2 \chi_{ij}^{(1)\top} \nabla^2 \beta_{(1)}^{,ij}
+\nabla^2 \chi_{ij}^{(1)\top} \beta_{(1)}^{,ij}
\;.\nonumber
\end{eqnarray}
Then, using the condition $\partial^i\chi^{(2)}_{{\scriptscriptstyle
\rm P}\ ij}=0$ and substituting $\beta^{(2)}$ 
we obtain an equation for  $d^{(2)}_i$: 
\begin{eqnarray}\label{eq:d2}
\nabla^2 d^{(2)}_i & = &-\frac{4}{3}\nabla^2\beta^{(2)}_{,i}
-\chi^{(2),j}_{{\scriptscriptstyle \rm S}\ ij}+
8\phi_{\scriptscriptstyle \rm S}^{(1),j}{\rm D}_{ij}\beta_{(1)} 
+\frac{16}{3}\phi_{\scriptscriptstyle \rm S}^{(1)}
\nabla^2\beta^{(1)}_{,i}
+\frac{2}{3}\nabla^2\beta_{(1)}^{,j}\beta^{(1)}_{,ij}
+\frac{10}{3}\beta_{(1)}^{,jk}\beta^{(1)}_{,ijk}
 \nonumber \\
&&
-\frac{8}{9}\nabla^2\beta_{(1)} \nabla^2\beta^{(1)}_{,i}
+2\beta_{(1)}^{,j\prime}{\rm D}_{ij}\beta_{(1)}^{\prime} 
+\frac{4}{3}\beta_{(1)}^{\prime}\nabla^2\beta^{(1)\prime}_{,i}
+\frac{4}{3}\beta_{(1)}^{,j}\nabla^2\beta^{(1)}_{,ij}
-4\chi_{ij,k}^{(1)\top} \beta^{(1),jk}
 \\
&&-2\left(\chi_{ij}^{(1)\top \prime}+2\frac{a'}{a}
\chi_{ij}^{(1)\top}\right)\beta^{(1)\prime,j}
-2\chi_{ij}^{(1)\top} \nabla^2 \beta_{(1)}^{,j}
-\frac{2}{3}\chi_{ik}^{(1)\top} \beta^{(1),jk}_{,i}
+\frac{4}{3}\chi_{jk,i}^{(1)\top} \beta^{(1),jk}
\nonumber.
\end{eqnarray}
Finally, using $\partial^i\omega_{{\scriptscriptstyle \rm P}\
i}^{(2)}=0$ and substituting $\beta^{(2)}$, we get an equation for
$\alpha^{(2)}$:
\begin{eqnarray}\label{eq:alpha2}
\nabla^2\alpha^{(2)} & = & \nabla^2\beta_{(2)}^{\prime}
-2\left(2\phi^{(1),i}_{\scriptscriptstyle \rm S}
+\beta_{(1)}^{\prime\prime,i} +\frac{1}{3}\nabla^2\beta_{(1)}^{,i}
\right)\beta^{(1)\prime}_{,i}
-2\beta_{(1)}^{,ij}\beta^{(1)\prime}_{,ij}
\nonumber \\
& &
-2\left(2\phi^{(1)}_{\scriptscriptstyle \rm S}
+\beta_{(1)}^{\prime\prime} -\frac{2}{3}\nabla^2\beta_{(1)}
\right)\nabla^2\beta_{(1)}^{\prime}
+2\chi_{ij}^{(1)\top}\beta^{(1)\prime,ij}
\;.
\end{eqnarray}
Having obtained, at least implicitly, all the parameters of the gauge
transformation to second order, one can in principle compute the metric
perturbations in the Poisson gauge from Eqs.\
(\ref{eq:psiP})--(\ref{eq:chiP}).

Similarly, once the parameters are known, the perturbations in any
scalar and 4-vector, and in particular those in the energy density and
in the 4-velocity of matter, follow trivially from Eqs.\
(\ref{eq:mut2})--(\ref{eq:vi2}).



\section{Evolution in the Poisson gauge}


\setcounter{equation}{0}

We have obtained in the previous section the general gauge
transformation to go from the synchronous to the Poisson gauge 
up to second order in metric perturbations. We can now apply it to 
the case of cosmological perturbations in a dust universe and 
compute  the perturbed metric in the Poisson gauge from the
solutions obtained in Section IV using the synchronous gauge.

\subsection{First-order perturbations} 

For the first order, replacing Eq.\ (\ref{eq:chis}) in Eq.\ (\ref{eq:v}) 
and using Eq.\ (\ref{eq:iii}), we obtain that the parameters of the
transformation are
\begin{eqnarray}
\alpha^{(1)}&=&\frac{\tau}{3} \varphi,\nonumber\\
\beta^{(1)}&=&\frac{\tau^2}{6} \varphi,
\end{eqnarray}
and $d^{(1)i}=0$, in the absence of vector modes in the initial 
conditions. 

For the metric perturbations we obtain from Eqs.\ (\ref{eq:i}),
(\ref{eq:iv}), (\ref{eq:ii}) and (\ref{eq:vii})
\begin{eqnarray}
\psi_{\scriptscriptstyle \rm P}^{(1)}& = 
&\phi_{\scriptscriptstyle \rm P}^{(1)}=\varphi,\nonumber\\
\chi_{{\scriptscriptstyle \rm P} ij}^{(1)}&=&\chi_{ij}^{\top(1)}.
\end{eqnarray}
These equations show the well-known result for scalar perturbations 
in the longitudinal gauge and the gauge invariance for tensor 
modes at the linear level.

The linear density contrast reads 
\begin{equation} 
\delta_{\scriptscriptstyle \rm P}^{(1)} = 
- 2 \varphi + {\tau^2 \over 6} \nabla^2 \varphi
\end{equation} 
while the first-order 4-velocity perturbation has components 
\begin{eqnarray}
v^{(1)0}_{\scriptscriptstyle \rm P} & = & - \varphi \\
v^{(1)i}_{\scriptscriptstyle \rm P} & = & - {\tau \over 3} \varphi^{,i} \;.
\end{eqnarray}

\subsection{Second-order perturbations} 

For the second-order parameters of the gauge transformation, replacing
the second-order perturbed metric obtained in Section IV in Eqs.\
(\ref{eq:beta2}) to (\ref{eq:alpha2}) we obtain
\begin{eqnarray}
\alpha^{(2)}&=&-\frac{2}{21} \tau^3 \Psi_0+\tau
\left(\frac{10}{9} \varphi^2 +4 \Theta_0\right)+\alpha^{(2)}_{({\rm t})},
\nonumber\\
\beta^{(2)}&=&\frac{\tau^4}{6}\left(\frac{1}{12}\varphi^{,i}\varphi_{,i}
-\frac{1}{7}\Psi_0\right)+\frac{\tau^2}{3}\left(\frac{7}{2} 
\varphi^2+6\Theta_0\right)+\beta^{(2)}_{({\rm t})},
\end{eqnarray}
with $\nabla^2 \Theta_0= \Psi_0-\frac{1}{3}\varphi^{,i}
\varphi_{,i}$ and
\begin{equation}
\label{eq:dii2}
\nabla^2 d^{(2)}_j = \frac{4\tau^2}{3} \left(- \varphi_{,j}
\nabla^2\varphi + \varphi^{,i}\varphi_{,ij}
- 2 \Psi_{0,j}\right)+\nabla^2 d^{(2)}_{({\rm t})j},
\end{equation}
where the quantities indicated by the subscript $({\rm t})$ stand for
the contributions arising from the presence of tensor modes at 
the linear level and are discussed in greater detail in
Appendix D.

For the perturbed metric we obtain
\begin{eqnarray}
\psi_{\scriptscriptstyle \rm P}^{(2)}& = 
&\tau^2\left(\frac{1}{6}\varphi^{,i}\varphi_{,i}
-\frac{10}{21}\Psi_0\right)+\frac{16}{3} \varphi^2+12  \Theta_0,
+\psi_{{\scriptscriptstyle \rm P}({\rm t})}^{(2)}
\nonumber\\
\phi_{\scriptscriptstyle \rm P}^{(2)}& = 
&\tau^2\left(\frac{1}{6}\varphi^{,i}\varphi_{,i}
-\frac{10}{21}\Psi_0\right)+\frac{4}{3} \varphi^2-8  \Theta_0,
+\phi_{{\scriptscriptstyle \rm P}({\rm t})}^{(2)}
\nonumber\\
\nabla^2 \omega_{\scriptscriptstyle \rm P}^{(2)i}& = 
&-\frac{8}{3}\tau\left(\varphi^{,i}
\nabla^2\varphi-\varphi^{,ij}\varphi_{,j}+2\Psi_0^{,i}\right)
+\nabla^2 \omega_{{\scriptscriptstyle \rm P}({\rm t})}^{(2)i}
\nonumber\\
\chi_{{\scriptscriptstyle \rm P} ij}^{(2)}& 
=&\tilde{\pi}_{ij}+\chi_{{\scriptscriptstyle \rm P}({\rm t}) ij}^{(2)}.
\label{eq:pg2}
\end{eqnarray}

The equations determining the contribution from linear tensor 
modes are given in Appendix D. Note that the contribution to 
$\psi_{\scriptscriptstyle \rm P}^{(2)}$ and 
$\phi_{\scriptscriptstyle \rm P}^{(2)}$ from linear scalar modes can be 
recovered, except for the sub-leading time-independent terms, 
by taking the weak-field limit of Einstein's theory (see, e.g., 
Ref.\cite{bi:pjep93}) and then expanding in powers of the perturbation 
amplitude. 

Also interesting is the way in which the second-order tensor modes,
generated by the non-linear growth of scalar perturbations, appear 
in this gauge: the transformation from the synchronous to the Poisson 
gauge has in fact dropped the Newtonian and post-Newtonian 
contributions, whose physical interpretation in terms of 
gravitational waves is highly non-trivial (see the discussion 
in Ref.\cite{bi:MT}); what remains is the tensor $\tilde{\pi}_{ij}$, whose 
evolution is governed by 
the equation 
\begin{equation}
\tilde{\pi}_{ij}''+\frac{4}{\tau}\tilde{\pi}_{ij}'-\nabla^2 \tilde{\pi}_{ij}=
- \frac{40}{3} {\cal T}_{ij} \;. 
\end{equation}
Its solution, Eq.\ (\ref{eq:tildepi}), 
contains a constant term, deriving from the vanishing initial conditions, 
plus a wave-like piece, having exactly the same form as linear tensor 
modes (cf. Eq.\ (\ref{eq:freegw})), whose 
amplitude is fixed by the source term ${\cal T}_{ij}$ (a quadratic 
combination of linear scalar modes). 
A more extended discussion of these tensor modes is given in 
Ref.\cite{bi:vale}. 

Finally, let us give the Poisson gauge expressions for the second-order 
density and 4-velocity perturbations. One has 
\begin{eqnarray} 
\delta^{(2)}_{\scriptscriptstyle \rm P} & = & {\tau^4\over 252} 
\biggl(5\bigl(\nabla^2 \varphi\bigr)^2 
+ 2 \varphi^{,ij}\varphi_{,ij} + 14 \varphi^{,i}\nabla^2\varphi_{,i} \biggr) + 
{\tau^2 \over 36} \biggl( - 21 \varphi^{,i} \varphi_{,i} + 24
\varphi\nabla^2 \varphi +{144\over 7} \Psi_0 - 6 \varphi^{,ij} 
\chi_{ij}^{(1)\top} \biggr) \nonumber
\\
&& 
+ {1 \over 4} 
\biggl( \chi^{(1)\top ij} \chi^{(1)\top}_{ij} - 
\chi^{(1)\top ij}_0 \chi^{(1)\top}_{0ij} \biggr) 
-{8 \over 3} \varphi^2 - 24 \Theta_0 
+ {3 \over 2} \phi^{(2)}_{{\scriptscriptstyle \rm S}({\rm t})}  - 
{6 \over \tau} \alpha^{(2)}_{({\rm t})} \;
\end{eqnarray}
and 
\begin{eqnarray}
v^{(2)0}_{\scriptscriptstyle \rm P} & = & \frac{\tau^2}{3} 
\left(-\frac{1}{6} \varphi^{,i}\varphi_{,i} + \frac{10}{7}
\Psi_0 \right) - \frac{7}{3}\varphi^2 - 12 \Theta_0 - 
\psi^{(2)}_{{\scriptscriptstyle \rm P}({\rm t})} \;,
\\
v^{(2)i}_{\scriptscriptstyle \rm P} & = & 
\frac{\tau^3}{9} \left( - \varphi^{,ij}\varphi_{,j} + \frac{6}{7} 
\Psi_0^{,i} \right) - 2 \tau \left( \frac{16}{9} \varphi\varphi^{,i} 
+ 2 \Theta_0^{,i} \right) 
- d^{(2)i\prime} - \beta^{(2)\prime ,i}_{({\rm t})} \;, 
\end{eqnarray}
with the vectors $d^{(2)i}$ defined in Eq.\ (\ref{eq:dii2}). 

In concluding this section, let us emphasize that all the second-order 
Poisson gauge expressions obtained here are new. Only a few terms in these 
expressions were already known in the literature, based on the 
weak-field limit of general relativity (e.g., Ref.\cite{bi:pjep93}). 


\section{Conclusions}

\setcounter{equation}{0}

In this paper we considered relativistic perturbations 
of a collisionless and irrotational fluid up to second order around the 
Einstein-de Sitter cosmological model. The most important 
phenomenon of second-order perturbation theory is mode mixing. 
An interesting consequence of this phenomenon is that primordial density 
fluctuations act as seeds for second-order gravitational waves. The specific 
form of these waves is gauge-dependent, as 
tensor modes are no longer gauge-invariant beyond the linear level. 
A second interesting effect is the generation of density fluctuations from
primordial tensor modes. One can even figure out a scenario in which 
no scalar perturbations were initially present, but they were later generated, 
as a second-order effect, by the non-linear evolution of a primordial 
gravitational-wave background. 

The first effect, which is discussed in some detail in Ref.\cite{bi:vale}, 
in the synchronous and comoving gauge also contains a term growing like 
$\tau^4$ and a second one growing like $\tau^2$: the first accounts 
for the Newtonian tidal induction of the  environment on the non-linear 
evolution of fluid elements, the second is a post-Newtonian tensor mode 
induced by the growth of the shear field. The remaining parts of this 
second-order tensor mode (excluding a constant term required by the 
vanishing initial conditions) oscillate with decaying amplitude inside
the horizon and describe true gravitational {\em waves}. 
Quite interesting is the fact that these are the only parts of
these second-order tensor modes which survive to the transformation leading to 
the Poisson gauge. 

The second effect is less known, and was only previously considered by 
Tomita back in the early 70's\cite{bi:tomita71-72}. 

One may naturally wonder whether there is any hope to detect the cosmological
stochastic gravitational-wave background produced at second order by scalar
fluctuations. It is, of course, the oscillating part of 
$\pi_{ij}$ which is relevant for earth or space detectors. The problem for 
these wave-like modes is that their energy density suffers the usual 
$a^{-4}$ dilution caused by free-streaming inside the Hubble radius, while at 
horizon-crossing their closure density is already extremely small,
$\Omega_{gw} \sim \delta_H^4$ (where $\delta_H$ is the {\em rms} density
contrast at horizon-crossing), because of their secondary origin.
More promising is the possibility that a non-negligible amount of
gravitational radiation can be produced during the strongly non-linear
stages of the collapse of proto-structures, an issue which would however 
require a fully non-perturbative approach. 

It should be stressed that, while many of our second-order terms had 
already been computed in the synchronous gauge, all our second-order 
Poisson-gauge expressions are new. 
This is a relevant result, as the latter gauge is the one which 
allows the easiest interpretation of the various physical effects. 
In particular, the second-order metric perturbations obtained by our 
method allow to compute self-consistently gravity-induced secondary 
anisotropies of the Cosmic Microwave Background. This calculation has been 
recently performed by Mollerach and Matarrese \cite{bi:mm97}, implementing a 
general scheme  introduced by Pyne and Carroll\cite{bi:pyne}. 

\acknowledgements

This work has been partially supported by the Italian MURST; MB thanks
SISSA for financial support, and S. Mollerach acknowledges the 
Vicerrectorado de investigaci\'on de la Universidad de Valencia
for financial support.




\appendix



\section{Taylor expansion of tensor fields} 


In this appendix we present some mathematical results used in Sec.\
II, concerning Taylor expansions of tensor fields on a manifold.
These results have already been presented in \cite{bi:P1}, where
analyticity of all relevant fields was assumed; they have been
generalized in \cite{bi:P2} to the case of $C^m$ fields. The theorems
obtained in \cite{bi:P1,bi:P2} are very general, concerning
perturbation theory at an arbitrary order $n$. In order to achieve
these general results it is very useful, or perhaps mandatory, to use
a fully geometrical approach.  However, for our purposes, it is useful
to summarize them in terms of coordinates and tensor components, as we
shall do in the following.  We assume that all quantities are as
smooth as necessary.


\subsection{One-parameter groups of transformations}


As disccused in Sec.\ II,
gauge choices for perturbations entail the comparison of the tensor
field representing a certain physical and/or geometrical quantity in
the perturbed spacetime with the tensor field representing the same
quantity in the background spacetime. Consequently, gauge transformations
entail the comparison of tensors at different points in the
background spacetime. A smallness parameter $\lambda$ is involved, so
that these comparisons are always carried out at the required order of
accuracy in $\lambda$, using Taylor
expansions\cite{parameter}. Differential geometry tells us that the
comparison of tensors is meaningful only when we consider them at the
same point. Therefore, supposing we want to compare a tensor field $T$ at
points $p$ and $q$, we need to define a transport law from $q$ to $p$.
This gives us two tensors at $p$, $T$ itself and the transported one,
which can be directly compared.

 The simplest transport law we need to consider is the Lie dragging by a
vector field, which allows us to compare $T$  with its pull-back
$\tilde{T}(\lambda)$ (the new tensor defined by this transport).  
To fix ideas, let us first consider, on a manifold $\cal M$, the comparison
of tensors at first order in $\lambda$ (which we shall define shortly). 
Suppose a coordinate system $x^\mu$ has been given on (an
open set of) $\cal M$, together with a vector field $\xi$. From ${\rm
d}x^\mu/{{\rm d}\lambda}=\xi^\mu$, $\xi$ generates on $\cal M$ a
congruence of curves $x^\mu(\lambda)$: thus $\lambda$ is the parameter
along the congruence.  Given a point $p$, this will always lie on one
of these curves, and we can always take $p$ to correspond to
$\lambda=0$ on this. The coordinates of a second point $q$ at a
parameter distance $\lambda$ from $p$ on the same curve, will be given
by
\begin{equation} 
\tilde{x}^\mu(\lambda)=x^\mu+\lambda\,\xi^\mu+\cdots\;,
\label{fopt}
\end{equation}
where the $x^\mu$ are the coordinates of $p$ and the $\tilde{x}^\mu$
are those of $q$, approximated here at first order in $\lambda$.  Eq.\
(\ref{fopt}) is usually called an `infinitesimal point
transformation', or an `active coordinate transformation' (see,
e.g. Ref. \cite{bi:raybook}, page 70; Ref. \cite{bi:stephani}, page
49; cf. also Ref. \cite{bi:weinberg}, page 291, and
\cite{bi:waldbook}, Appendix C). At the same time, we may think that a
new coordinate system $y^\mu$ has been introduced on ${\cal M}$, 
{\it defined} in such a
way that the $y$-coordinates of the point $q$ coincide with the
$x$-coordinates of the point $p$; using (\ref{fopt}) it then follows
from this definition that
\begin{eqnarray}
y^\mu(q):=x^\mu(p)&=&x^\mu(q) -\lambda \xi^\mu(x(p)) +\cdots \nonumber
\\
& \simeq & x^\mu(q) -\lambda \xi^\mu(x(q)) +\cdots\;.
\label{fopct}
\end{eqnarray}
In practice, we have in this way defined at every point a
`passive coordinate transformation' (i.e., just an ordinary
relabeling of point's names), that at first order reads:
\begin{equation}
y^\mu(\lambda)=x^\mu -\lambda \xi^\mu +\cdots\;.
\label{fopct'}
\end{equation}

Suppose now that a tensor field has been given on $\cal M$; e.g., to fix 
ideas, consider the vector field $Z$ with components $Z^\mu$ in the
$x$-coordinate system. In the same way that we defined a new coordinate
system $y^\mu$ once a relation between points was assigned through
(\ref{fopt}) by the action of $\xi$, so 
we can now define a new vector field $\tilde{Z}$, with components 
$\tilde{Z}^\mu$ in the $x$-coordinates, such that these components at the
coordinate point $x^\mu(p)$ are equal to the components
$Z^{\prime\mu}$  the old
vector $Z$ has in the $y$-coordinates at the coordinate point $y(q)$:
\begin{equation}
\tilde{Z}^\mu(x(p)):=Z^{\prime\mu}(y(q))
=\left[\frac{\partial
y^\mu}{\partial x^\nu}\right]_{x(q)}Z^\nu(x(q))\; .
\label{eq:defZ}
\end{equation}
The last equality in this equation is just the ordinary (passive)
transformation between the components of $Z$ in the two coordinate
systems: we need it in order to relate $\tilde{Z}$ and $Z$ in a single
system (the $x$-frame here), thus eventually obtaining  a
covariant relation. Indeed,
substitution of (\ref{fopct'}) into (\ref{eq:defZ}) and a first order
expansion in $\lambda$ about $x(p)$ in the RHS gives
\begin{eqnarray}
\tilde{Z}^\mu(\lambda)& 
=& Z^\mu+\lambda\pounds_\xi Z^\mu +\cdots\;, \label{fopb}\\
\pounds_\xi Z^\mu&:=& Z^\mu_{,\nu}\xi^\nu -\xi^\mu_{,\nu}Z^\nu\;,\label{lddef}
\end{eqnarray}
where, given that the point $p$ is arbitrary, the dependence of all
terms by $x(p)$ has been omitted.
The vector field $\tilde{Z}$ is called the pull-back of $Z$, because
is defined by dragging $Z$ back from $q$ to $p$, an operation that
gives at $p$ a new vector with components $\tilde{Z}^\mu$, given by
(\ref{eq:defZ}). In the particular case of the transformation
(\ref{fopct'}) this is the Lie-dragging.  Now, having at the same
point two vectors, these can be directly compared: at first order,
$\tilde{Z}(\lambda)$ and $Z$ are related by (\ref{fopb}),
(\ref{lddef}).  In fact, 
in the limit $\lambda \to 0$, is  this comparison that allows us
to define the Lie derivative, with components (\ref{lddef});
Eq.\ (\ref{eq:lieder}) below generalizes this to a generic tensor $T$.

Although the story so far is a textbook one (cf.
\cite{bi:waldbook,bi:raybook,bi:stephani,bi:weinberg}), recalling it
in some detail 
 allows us to easily extend it to higher order.
First, one has to realize that (\ref{fopt}) is just the first order
Taylor expansion about $x(p)$ of the solution of the ordinary
differential equation ${\rm d}x^\mu/{{\rm d}\lambda}=\xi^\mu$
 defining the congruence $x^\mu(\lambda)$ associated with $\xi$.
The exact solution of this
equation is the Taylor series (cf., e.g., \cite{schutzbook}, page 43)
\begin{equation}
x^\mu (q)=x^\mu(p) +\lambda \xi^\mu (x(p))
+\frac{\lambda^2}{2}\, {\xi^\mu}_{,\nu} \xi^\nu (x(p)) +\cdots\;,
\label{lemma1coord1}
\end{equation}
on using $dx^\mu/d\lambda= \xi^\mu$, $ {\rm
d}^2x^\mu/ {\rm d}\lambda^2 =\xi^\mu_{,\nu}\xi^\nu$, etc.
In practice, since $p$ and $q$ are arbitrary, we may simply write
\begin{eqnarray} 
\tilde{x}^\mu(\lambda)&=&x^\mu+\lambda\,\xi^\mu+{\lambda^2\over
2}\,{\xi^\mu}_{,\nu}\xi^\nu+\cdots\;,
\label{lemma1coord} \\
&=& \exp[\lambda\pounds_\xi]x^\mu\;. 
\label{eq:expxmu}
\end{eqnarray}
The latter exponential notation is useful, in that it allows us  to see
the coordinate functions $\tilde{x}^\mu$ as the
pull-backs of the functions $x^\mu$ given by the exponential pull-back operator
$\exp [\lambda\pounds_\xi]$. Furthermore, it is clearly seen by 
 $\exp [(\lambda_1+\lambda_2)\pounds_\xi]=\exp [\lambda_1\pounds_\xi]\, 
\exp[\lambda_2\pounds_\xi]$ that the point transformations
(\ref{lemma1coord})
form a one-parameter group of transformations.
Using again the definition $y^\mu(q):=x^\mu(p)$ for the
$y$-coordinates, we get from (\ref{lemma1coord1})
\begin{equation}
y^\mu(\lambda)=x^\mu-\lambda\,\xi^\mu+{\lambda^2\over
2}\,{\xi^\mu}_{,\nu} \xi^\nu+\cdots\;,
\label{lemma2coord'''}
\end{equation}
 on expanding all  terms about $x(q)$, eventually  omitting again the
$x(q)$ dependence, since $q$ is arbitrary.
Finally, using (\ref{lemma2coord'''}) into (\ref{eq:defZ}) and
expanding all the terms about $x(p)$, we get the $x$-components
$\tilde{Z}^\mu(\lambda)$ of the pull-back $\tilde{Z}(\lambda)$, which 
reads\cite{notation}:
\begin{eqnarray}
\tilde{Z}^\mu(\lambda) & =& [\,\exp[\lambda\pounds_\xi]\, Z\,]^\mu 
\label{eq:expZ} \\
&=&
Z^\mu+\lambda\pounds_\xi Z^\mu
+\frac{\lambda^2}{2}\pounds^2_\xi
Z^\mu +\cdots\;.
\label{eq:exp1}
\end{eqnarray}

Eq.\ (\ref{eq:defZ}) is readily generalized to  more general
tensors than $Z$: we simply have to add to the RHS of   (\ref{eq:defZ})
the right number of transformation matrices. Thus, the
pull-back $\tilde{T}$ of  a tensor field $T$ of type (p,q) is defined
by having $x$-components given by 
\begin{eqnarray}
\lefteqn{\tilde{T}^{\mu_1\cdots\mu_p}{}_{\nu_1\cdots\nu_q}(x(p)):=
 T^{\prime \mu_1\cdots\mu_p}{}_{\nu_1\cdots\nu_q}(y(q)) 
}\hspace{1.25truecm} \nonumber \\
&=&
\left[\, \frac{ \partial y^{\mu_1} }{\partial x^{\alpha_1} } 
\cdots
 \frac{ \partial y^{\mu_p} }{\partial x^{\alpha_p} }
\, \frac{ \partial x^{\beta_1} }{\partial y^{\nu_1} } 
\cdots
 \frac{ \partial x^{\beta_q} }{\partial y^{\nu_q} }\, \right]_{x(q)} 
T^{\alpha_1\cdots\alpha_p}{}_{\beta_1\cdots\beta_q}(x(q))\; .
\label{eq:Tpb1}
\end{eqnarray}
Using (\ref{lemma2coord'''}) as above then gives, omitting indices for
brevity, 
\begin{equation}
\tilde{T}(\lambda)=T+\lambda \pounds_{\xi} T
+\frac{\lambda^2}{2}\pounds^2_{\xi} T + \cdots \;.
\label{tildet}
\end{equation}

To summarize, each of the diffeomorphisms forming  a  one-parameter group, as
mathematicians call the transformations generated by a vector field
$\xi$ and  represented in coordinates by 
(\ref{eq:expxmu}), gives rise to  a new
field, the pull-back $\tilde{T}(\lambda)$, from any given tensor field
$T$ and for any given value of $\lambda$.
Thus $\tilde{T}(\lambda)$ and $T$ may be compared at every point, 
which allows one to define the Lie derivative along $\xi$   
as the limit $\lambda \to
0$ of the difference
$\tilde{T}(\lambda) -T$:
\begin{equation}
\pounds_\xi T :=
 \left[ \frac{{\rm d}~~}{{\rm d}\lambda}\right]_{\lambda=0}  
\tilde{T}(\lambda)=\lim_{\lambda\rightarrow 0}\frac{1}{\lambda}\,
\left(\tilde{T}(\lambda)-T\right)\;.
\label{eq:lieder}
\end{equation}
At higher order we have
\begin{equation}
\pounds_\xi^k T:=\left[\frac{{\rm d}^k~~}{{\rm d} 
\lambda^k}\right]_{\lambda=0}  \tilde{T}(\lambda)\;.
\label{??}
\end{equation}
On the other hand, 
the relation at each point  between any tensor field $T$ and its
pull-back $\tilde{T}(\lambda)$ is expressed at the required order of accuracy
by the Taylor expansion  (\ref{tildet}).


\subsection{One-parameter families of transformations}


In order to proceed, considering more general point transformations
than (\ref{lemma1coord}) and more general Taylor expansions that
(\ref{tildet}), some general remarks are in order.  First, it should
be noticed that the definition $y^\mu(q):=x^\mu(p)$ for the
$y$-coordinate system is completely general, given a first coordinate
system (the $x$-frame here) and any suitable association between pairs
of points (more precisely, any diffeomorphism), of which the
one-parameter group of transformations (\ref{lemma1coord}) is a
particular example. Second, the same generality is present in the
definition of the pull-back, Eq.\ (\ref{eq:Tpb1}), which is also
independent from the specific type of transformation chosen.

As we said  in Section II, exact gauge transformations  do
not form a one-parameter group, but a one-parameter family\cite{bi:P1,bi:P2}.
However the consequences of this fact show up only with non linearity,
which is why at first order gauge transformations are approximated by
(\ref{fopt}), (\ref{fopct'})
(cf. \cite{bi:waldbook,bi:raybook,bi:stephani,bi:weinberg}). 
Therefore, having in mind a second order treatment of perturbations, 
the question we now have to deal with is twofold: {\it i)} which is
the general form of families of transformations that depend on one
parameter (one-parameter families of diffeomorphisms), 
 but do not form a group; {\it ii)} which is the form of the Taylor
expansion of the pull-back $\tilde{T}(\lambda)$ of a tensor $T$ generated by
one such one-parameter family of transformations.

In \cite{bi:P1,bi:P2} (cf. also \cite{bi:GL})
 we have shown that the action of any given one-parameter
family of  transformations can be represented by the successive action
of one-parameter groups, in a fashion that, to order $\lambda^2$,
 reminds us the motion of the knight on the chess-board:
\begin{equation}
 \tilde{x}^\mu(\lambda)=x^\mu
+\lambda\,\xi_{(1)}^\mu+{\lambda^2\over
2}\,\left({\xi_{(1)}^\mu}_{,\nu}\xi_{(1)}^\nu
+\xi_{(2)}^\mu\right)+\cdots\;.
\label{lemma2coord}
\end{equation}
A vector field $\xi_{(k)}$ is associated to  the $k$-th one-parameter group 
of transformations, with parameter $\lambda_k$ (we denote
 $\lambda_1=\lambda$). 
Similarly to the knight, the action  
of the transformation   (\ref{lemma2coord}) first
moves from point $p$ (with coordinates $x^\mu$) by an amount $\lambda$
along the integral
curve of  $\xi_{(1)}$ [i.e.,
according to Eq.\ (\ref{lemma1coord})]; then, it moves along the
integral curve of $\xi_{(2)}$ by an amount $\lambda_2=\lambda^2/2$. At
each $k$-th higher order, a new vector field $\xi_{(k)}$ is involved,
generating a motion by $\lambda_k=\lambda^k/k!$. Thus, the action of a
one-parameter family of transformations is approximated, at order $k$,
by a `knight transformation' of order $k$ (see \cite{bi:P1}, Theorem
2), of which (\ref{lemma2coord}) is the second order example.

Given the `knight transformation' (\ref{lemma2coord}), we can now
use it to define the $y$-coordinates, which will be given by
\begin{equation}
y^\mu(q):= x^\mu(p)=x^\mu(q)-\lambda\,\xi_{(1)}^\mu(x(p))
-\frac{\lambda^2}{2}\,\left({\xi_{(1)}^\mu}_{,\nu}(x(p))\
 \xi_{(1)}^\nu(x(p))+\xi_{(2)}^\mu(x(p))\right)+\cdots\;.
\label{gosh}
\end{equation}
Expanding the various quantities on the RHS around $q$, and omitting
the $x(q)$ dependence, 
(\ref{gosh}) becomes finally
\begin{equation}
 y^\mu(\lambda)=x^\mu-\lambda\,\xi_{(1)}^\mu+\frac{\lambda^2}{2}
\,\left({\xi_{(1)}^\mu}_{,\nu}\ 
\xi_{(1)}^\nu -\xi_{(2)}^\mu \right)+
\cdots\;.
\label{lemma2coord''}
\end{equation}
Using again the case of the vector field $Z$ as our paradigmatic
example, we can now derive the pull-back $\tilde{Z}(\lambda)$ generated by a
one-parameter family of transformations. 
Substituting (\ref{lemma2coord''}) into (\ref{eq:defZ}), and expanding
again every term about $x(p)$, we obtain the $x$-components
$\tilde{Z}^\mu(\lambda)$ of $\tilde{Z}(\lambda)$,
 which (after properly collecting terms) 
at second order read
\begin{equation}
\tilde{Z}^\mu(\lambda)=Z^\mu+\lambda \pounds_{\xi_{(1)}} Z^\mu
+\frac{\lambda^2}{2}\left(\pounds^2_{\xi_{(1)}}
+\pounds_{\xi_{(2)}}\right)Z^\mu + \cdots\;.
\label{lemma2explicZ}
\end{equation}
For a generic tensor $T$, again omitting indices for brevity, use of
(\ref{lemma2coord''}) into (\ref{eq:Tpb1})
 obviously gives
\begin{equation}
\tilde{T}=T+\lambda \pounds_{\xi_{(1)}} T
+\frac{\lambda^2}{2}\left(\pounds^2_{\xi_{(1)}}
+\pounds_{\xi_{(2)}}\right)T + \cdots \;.
\label{lemma2explic}
\end{equation}

\section{ Second-order perturbations of useful quantities in the synchronous 
gauge} 

In this appendix we report the expansions up to second order of a number of 
tensors, which have been used in deriving the results of Section IV.  
All calculations are performed in the synchronous and comoving gauge,
assuming an Einstein-de Sitter background. No subscripts 
$\scriptscriptstyle \rm S$ on
synchronous-gauge quantities will be used in this appendix. 

The covariant conformal spatial metric tensor is expanded as follows, 
\begin{equation}
\gamma_{ij} = \delta_{ij} + \gamma_{ij}^{(1)} + {1 \over 2} 
\gamma_{ij}^{(2)} \;. 
\end{equation} 
The corresponding contravariant metric takes the form
\begin{equation}
\gamma^{ij} = \delta^{ij} - \gamma^{(1)ij} - {1 \over 2} 
\gamma^{(2)ij} + \gamma^{(1)ik} \gamma^{(1)j}_k \;,
\end{equation} 
where the indices of the perturbations $\gamma_{ij}^{(1,2)}$ are raised by 
$\delta^{ij}$. 

The extrinsic curvature tensor $\vartheta^i_{~j}$ up to second order reads 
\begin{equation}
\vartheta^i_{~j} = {1 \over 2} \biggl( {g^{(1)i}_{~j}}' + {1 \over 2} 
{\gamma^{(2)i}_{~j}}' - \gamma^{(1)ik} {\gamma^{(1)}_{kj}}' \biggr) \;. 
\end{equation} 

The square root of the metric determinant is 
\begin{equation}
\gamma^{1/2} = 1 + {1 \over 2} \gamma^{(1)i}_{~i} + {1 \over 4} 
\gamma^{(2)i}_{~i} 
+ {1 \over 8} \bigl(\gamma^{(1)i}_{~i}\bigr)^2 - {1 \over 4} \gamma^{(1)ij} 
\gamma^{(1)}_{ij} \;,
\end{equation} 
with inverse 
\begin{equation}
\gamma^{-1/2} = 1 - {1 \over 2} \gamma^{(1)i}_{~i} - {1 \over 4} 
\gamma^{(2)i}_{~i} 
+ {1 \over 8} \bigl(\gamma^{(1)i}_{~i}\bigr)^2 + {1 \over 4} \gamma^{(1)ij} 
\gamma^{(1)}_{ij}\;. 
\end{equation} 

>From these quantities we can easily get the density contrast
\begin{eqnarray}
\delta &=& - {1 \over 2} \gamma^{(1)i}_{~i} + {1 \over 2} \gamma^{(1)i}_{0i} 
+ \delta_0 - {1 \over 4} \gamma^{(2)i}_{~i} 
+ {1 \over 8} \bigl(\gamma^{(1)i}_{~i}\bigr)^2 + {1 \over 8} 
\bigl(\gamma^{(1)i}_{0i}\bigr)^2 
- {1 \over 4} \gamma^{(1)i}_{~i} \gamma^{(1)j}_{0j} + 
\nonumber\\
&& 
{1 \over 4} \gamma^{(1)ij} \gamma^{(1)}_{ij} - {1 \over 4} 
\gamma^{(1)ij}_0 \gamma^{(1)}_{0ij} 
- {1 \over 2} \gamma^{(1)i}_{i} \delta_0 
+ {1 \over 2} \gamma^{(1)i}_{0i} \delta_0 \;, 
\end{eqnarray} 
having assumed as initial conditions $\gamma^{(2)}_{0ij} = 0$ and 
$\delta^{(2)}_0=0$ (i.e. $\delta_0=\delta^{(1)}_0$). 

The Christoffel symbols up to second order read 
\begin{equation}
\Gamma^i_{jk} = {1 \over 2} \biggl( \gamma^{(1)i}_{~j~,k} + 
\gamma^{(1)i}_{~k~,j} -
\gamma^{(1),i}_{jk} \biggr) + {1 \over 4} 
\biggl( \gamma^{(2)i}_{~j~,k} + \gamma^{(2)i}_{~k~,j} - 
\gamma^{(2),i}_{jk} \biggr) - 
{1 \over 2} \gamma^{(1)i}_{~\ell} \biggl( \gamma^{(1)\ell}_{~j~,k} + 
\gamma^{(1)\ell}_{~k~,j} 
- \gamma^{(1),\ell}_{jk} \biggr) \;, 
\end{equation} 
from which, after a lengthy but straightforward calculation, the conformal 
Ricci tensor of the spatial hypersurface, 
\begin{eqnarray}
{\cal R}^i_{~j} & = & {1 \over 2} \biggl( \gamma^{(1)ik}_{~~~~,jk} + 
\gamma^{(1)k~,i}_{~j~,k} - \nabla^2 \gamma^{(1)i}_{~j} - 
\gamma^{(1)k~,i}_{~k~,j} 
\biggr) + {1 \over 4} \biggl( \gamma^{(2)ik}_{~~~~,jk} + 
\gamma^{(2)k~,i}_{~j~,k} - \nabla^2 \gamma^{(2)i}_{~j} - 
\gamma^{(2)k~,i}_{~k~,j} \biggr) 
\nonumber\\
&& + {1 \over 2} \biggl[ \gamma^{(1)ik} \biggl( \nabla^2 \gamma^{(1)}_{kj} 
+ \gamma^{(1)\ell}_{~\ell~,jk} - \gamma^{(1)\ell}_{~k~,\ell j} - 
\gamma^{(1)\ell}_{~j~,\ell k} \biggr) + 
\gamma^{(1)\ell k} \biggl( \gamma^{(1)i}_{~j~,k\ell} + 
\gamma^{(1),i}_{~\ell k~,j} -
\gamma^{(1)i}_{~k~,j\ell} - \gamma^{(1),i}_{~j\ell~,k} \biggr) 
\nonumber\\
&& + \gamma^{(1)\ell k}_{~~~~~,\ell} \biggl( \gamma^{(1)i}_{~j~,k} 
+ \gamma^{(1)i}_{~k~,j} - \gamma^{(1),i}_{jk} \biggr) 
+ \gamma^{(1)\ell i}_{~~~~,k} \gamma^{(1),k}_{~\ell j} 
- \gamma^{(1)\ell i}_{~~~~,k} \gamma^{(1)k}_{~j~,\ell} 
+ {1 \over 2} \gamma^{(1)\ell k}_{~~~~,j} \gamma^{(1),i}_{~\ell k} 
\nonumber\\
&& 
+ {1 \over 2} \gamma^{(1)m}_{~m~,\ell} \biggl( \gamma^{(1)\ell i}_{~~~~,j} + 
\gamma^{(1)\ell,i}_{~j} - \gamma^{(1)i,\ell}_{~j} \biggr) \biggr] 
\;, 
\end{eqnarray} 
and its trace 
\begin{eqnarray}
{\cal R}  & = & \gamma^{(1)\ell k}_{~~~~~,\ell k} - 
\gamma^{(1)k~,\ell}_{~k~~,\ell} 
+ {1 \over 2} \biggl( \gamma^{(2)\ell k}_{~~~~~,\ell k} - 
\gamma^{(2)k~,\ell}_{~k~~,\ell} \biggr) + \gamma^{(1)jk} 
\biggl( \nabla^2 \gamma^{(1)}_{jk} 
+ \gamma^{(1)\ell}_{~\ell~,jk} - 2 \gamma^{(1)\ell}_{~j~,\ell k} \biggr) 
\nonumber\\
&& 
+ \gamma^{(1)\ell k}_{~~~~,\ell} 
\biggl( \gamma^{(1)j}_{~j~,k} - \gamma^{(1),j}_{jk} \biggr) 
+ {3 \over 4} \gamma^{(1)\ell j}_{~~~~,k} \gamma^{(1),k}_{~\ell j} 
- {1 \over 2} \gamma^{(1)\ell j}_{~~~~,k} \gamma^{(1)k}_{~j~,\ell} 
- {1 \over 4} \gamma^{(1)j,\ell}_{~j} \gamma^{(1)k}_{~k~,\ell} 
\; 
\end{eqnarray} 
follow.

\section{ Second-order perturbations generated by linear tensor modes} 

In this appendix we report the equations which allow to determine 
the second-order perturbations which arise due to the presence of tensor 
modes at the linear level: these are originated both by the 
coupling of scalar and tensor modes and by tensor-tensor mode couplings. 
All calculations are performed in the synchronous and comoving gauge,
assuming an Einstein-de Sitter background. 

The equations which follow refer only to those parts of the second-order 
metric perturbations which involve tensor modes in the source terms
(hence the subscript $({\rm t})$). 
\\

\noindent
{\bf Raychaudhuri equation} 

\begin{eqnarray} 
{\phi^{(2)}_{{\scriptscriptstyle \rm S}({\rm t})}}'' && + {2 \over \tau} 
{\phi^{(2)}_{{\scriptscriptstyle \rm S}({\rm t})}}' - {6 \over \tau^2} 
\phi^{(2)}_{{\scriptscriptstyle \rm S}({\rm t})} 
= {\tau^2 \over 9} \varphi^{,ij} \nabla^2 
\chi^{(1)\top}_{ij} - {1 \over 6} 
{\chi^{(1)\top ij}}' {\chi^{(1)\top}_{ij}}' + 
{2 \over 3 \tau} {\chi^{(1)\top ij}}' \chi^{(1)\top}_{ij} 
\nonumber\\ 
&& 
- {1 \over 3} \chi^{(1)\top ij} \nabla^2 \chi^{(1)\top}_{ij} 
+ {1 \over \tau^2} \biggl( \chi^{(1)\top ij} \chi^{(1)\top}_{ij}  -
\chi^{(1)\top ij}_0 \chi^{(1)\top}_{0ij} \biggr) \equiv 
{\cal Q}({\bf x},\tau)
\;; \label{eq:ray}
\end{eqnarray} 

\noindent
{\bf energy constraint} 

\begin{eqnarray}
{2 \over \tau} {\phi^{(2)}_{{\scriptscriptstyle \rm S}({\rm t})}}' 
&& - {1 \over 3} 
\nabla^2 \phi^{(2)}_{{\scriptscriptstyle \rm S}({\rm t})} +{6 \over \tau^2} 
\phi^{(2)}_{{\scriptscriptstyle \rm S}({\rm t})} - 
{1 \over 12} \chi^{(2)ij}_{{\scriptscriptstyle \rm S}({\rm t}),ij} 
\nonumber\\ 
&& = {5 \tau \over 18} {\chi^{(1)\top ij}}' \varphi_{,ij} + 
{5 \over 9} \chi^{(1)\top ij} \varphi_{,ij} - {\tau^2 \over 18} 
\nabla^2 \chi^{(1)\top ij} \varphi_{,ij} - {\tau^2 \over 36} 
\chi^{(1)\top ij,k} \varphi_{,ijk} 
\nonumber\\ 
&&
- {1 \over 24} 
{\chi^{(1)\top ij}}' {\chi^{(1)\top}_{ij}}' 
- {2 \over 3\tau} {\chi^{(1)\top ij}}' \chi^{(1)\top}_{ij}
+ {1 \over 6} \chi^{(1)\top ij} \nabla^2 \chi^{(1)\top}_{ij}
\nonumber\\ 
&& 
+ {1 \over 8} \chi^{(1)\top ij,k} \chi^{(1)\top}_{ij,k}
- {1 \over 12} \chi^{(1)\top ij,k} \chi^{(1)\top}_{kj,i}
- {1 \over \tau^2} \biggl( \chi^{(1)\top ij} \chi^{(1)\top}_{ij} 
- \chi^{(1)\top ij}_0 \chi^{(1)\top}_{0ij} \biggr) 
\;; \label{eq:ec}
\end{eqnarray} 
\newpage
\noindent
{\bf momentum constraint} 

\begin{eqnarray}
2 {\phi^{(2)}_{{\scriptscriptstyle \rm S}({\rm t}),j}}' && + {1 \over 2} 
{\chi^{(2)i}_{{\scriptscriptstyle \rm S}({\rm t})j,i}}' 
= {\tau^2 \over 3} \biggl[ \biggl( {\chi^{(1)\top ik}_{~~~~~~~,j}}' - 
{\chi^{(1)\top k,i}_{~~~~j}}' \biggr) \varphi_{,ik} + 
{1 \over 2} {\chi^{(1)\top ik}}' \varphi_{,ikj} 
- {1 \over 2} {\chi^{(1)\top k}_{~~~j}}' \nabla^2 \varphi_{,k} \biggr]  
\nonumber\\ 
&& 
+ {\tau \over 3} \chi^{(1)\top ik}_{~~~~~~~,j} \varphi_{,ik} 
+ {5 \over 3} {\chi^{(1)\top i}_{~~~j}}' \varphi_{,i} + \chi^{(1)\top ik} 
\biggl( {\chi^{(1)\top}_{kj,i}}' - {\chi^{(1)\top}_{ki,j}}' 
\biggr) - {1 \over 2} \chi^{(1)\top ik}_{~~~~~~~,j} {\chi^{(1)\top}_{ik}}' 
\;;\label{eq:mc}
\end{eqnarray} 

\noindent
{\bf evolution equation} 

\begin{eqnarray}
- \biggl({\phi^{(2)}_{{\scriptscriptstyle \rm S}({\rm t})}}'' 
&& + {4 \over \tau} 
{\phi^{(2)}_{{\scriptscriptstyle \rm S}({\rm t})}}' \biggr) \delta^i_{~j} + 
{1 \over 2} \biggl( {\chi^{(2)i}_{{\scriptscriptstyle \rm S}({\rm t})j}}'' + 
{4 \over \tau} {\chi^{(2)i}_{{\scriptscriptstyle \rm S}({\rm t})j}}' \biggr) + 
\phi^{(2),i}_{{\scriptscriptstyle \rm S}({\rm t}),j} 
- {1 \over 4} \chi^{(2)k\ell}_{{\scriptscriptstyle \rm S}({\rm t}),k\ell} 
\delta^i_{~j} 
+ {1 \over 2} \chi^{(2)ki}_{{\scriptscriptstyle \rm S}({\rm t}),kj} 
\nonumber\\ 
&& + {1 \over 2} \chi^{(2)k~~~~~,i}_{{\scriptscriptstyle \rm S}({\rm t})j,k} 
- {1 \over 2} 
\nabla^2 \chi^{(2)i}_{{\scriptscriptstyle \rm S}({\rm t})j} = 
- {2 \tau \over 3} 
{\chi^{(1)\top}_{kj}}' \varphi^{,ik} - {2 \tau \over 3} 
{\chi^{(1)\top ik}}' \varphi_{,kj} + {\tau \over 3} 
{\chi^{(1)\top i}_{~~j}} \nabla^2 \varphi 
+ {\tau \over 6} {\chi^{(1)\top k\ell}}' \varphi_{,k\ell} \delta^i_{~j} 
\nonumber\\ 
&& 
+ {10 \over 3} \chi^{(1)\top i}_{~~j} \nabla^2 \varphi + 
{25 \over 3}  \chi^{(1)\top k\ell} \varphi_{,k\ell} \delta^i_{~j} 
- {10 \over 3} \chi^{(1)\top ik} \varphi_{,jk} 
- {10 \over 3} \chi^{(1)\top}_{jk} \varphi^{,ik} 
+ {10 \over 3} \nabla^2 \chi^{(1)\top i}_{~~~j} \varphi
\nonumber\\ 
&& 
+ {\tau^2 \over 3} \biggl( \chi^{(1)\top i,k\ell}_{~~~j} + 
\chi^{(1)\top k\ell,i}_{~~~~~~~~~,j} - \chi^{(1)\top ki,\ell}_{~~~~~~~~~,j} 
- \chi^{(1)\top k,i\ell}_{~~~j} \biggr) \varphi_{,k\ell}
\nonumber\\ 
&& 
+ \biggl( \chi^{(1)\top i,k}_{~~~j} - \chi^{(1)\top ik}_{~~~~~~,j} 
- \chi^{(1)\top k,i}_{~~~j} \biggr) 
\biggl( {5 \over 3} \varphi_{,k} + {\tau^2 \over 6} \nabla^2 \varphi_{,k} 
\biggr) 
\nonumber\\ 
&& + {\tau^2 \over 6} \biggl( \chi^{(1)\top k\ell,i} \varphi_{,k\ell j}
+ \chi^{(1)\top k\ell}_{~~~~~~,j} \varphi_{,k\ell i} 
- \nabla^2 \chi^{(1)\top k\ell} \varphi_{,k\ell} \delta^i_{~j} 
\biggr) - {\tau^2 \over 12} \chi^{(1)\top k\ell,m} \varphi_{,k\ell m} 
\delta^i_{~j} + {\chi^{(1)\top ik}}' {\chi^{(1)\top}_{kj}}' 
\nonumber\\ 
&& - {1 \over 8} {\chi^{(1)\top k\ell}}' {\chi^{(1)\top}_{k\ell}}' 
\delta^i_{~j} + \chi^{(1)\top k\ell} \biggl( \chi^{(1)\top i}_{~~~k~~,j\ell} 
+ \chi^{(1)\top ,i}_{~~kj~~,\ell} 
- \chi^{(1)\top i}_{~~~j~~,k\ell} - \chi^{(1)\top ,i}_{~~k\ell~~,j} \biggr) 
\nonumber\\ 
&& 
+ \chi^{(1)\top ki}_{~~~~~~,\ell} \biggl(\chi^{(1)\top \ell}_{~~j~~,k} 
- \chi^{(1)\top ,\ell}_{~~kj} \biggr) 
- {1 \over 2} \chi^{(1)\top k\ell}_{~~~~~~,j} \chi^{(1)\top ,i}_{~~k\ell} 
+ {1 \over 2} \chi^{(1)\top k\ell} \nabla^2 \chi^{(1)\top}_{~k\ell} 
\delta^i_{~j}
\nonumber\\ 
&&  + {3 \over 8} \chi^{(1)\top km}_{~~~~~~~~,\ell} 
\chi^{(1)\top,\ell}_{~km} \delta^i_{~j} 
- {1 \over 4} \chi^{(1)\top km}_{~~~~~~~,\ell} 
\chi^{(1)\top\ell}_{~m~~,k} \delta^i_{~j} 
\;. \label{eq:ee}
\end{eqnarray} 

The Raychaudhuri equation can be easily solved for 
$\phi^{(2)}_{{\scriptscriptstyle \rm S}({\rm t})}$ 
by means of the Green method. The
resulting expression has been given in Eq.\ (\ref{eq:qtau}) of the main text. 
By replacing it in the remaining equations one can in principle obtain 
the traceless tensor $\chi^{(2)}_{{\scriptscriptstyle \rm S}({\rm t})ij}$ 
by integration.

\section{Tensor contribution to the second-order gauge transformation} 

In this appendix we show how to compute the contribution from linear 
tensor modes to the perturbed metric in the Poisson gauge by performing 
a gauge transformation from the synchronous gauge perturbed metric 
obtained from Appendix C. The equations for the gauge transformation
parameters involved are obtained straightforwardly from Eqs.\
(\ref{eq:beta2}) to (\ref{eq:alpha2})
\begin{eqnarray}\label{eq:beta2t}
\nabla^2\nabla^2\beta^{(2)}_{({\rm t})} & = & -\frac{3}{4}
\chi^{(2)ij}_{{\scriptscriptstyle \rm S}({\rm t}), ij}
-\frac{\tau}{2}\left(\chi_{ij}^{(1)\top \prime}+\frac{4}{\tau}
\chi_{ij}^{(1)\top}\right)\varphi^{,ij}
-\frac{5\tau^2}{12}\chi_{ij,k}^{(1)\top}\varphi^{,ijk}
- \frac{\tau^2}{3}\chi_{ij}^{(1)\top} \nabla^2 \varphi^{,ij}
+\frac{\tau^2}{6}\nabla^2 \chi_{ij}^{(1)\top} \varphi^{,ij}
\;,\nonumber \\
\nabla^2\alpha^{(2)}_{({\rm t})}& = & \nabla^2\beta^{(2)\prime}_{({\rm t})} 
+\frac{2\tau}{3} \chi_{ij}^{(1)\top} \varphi^{,ij}\;,
\\
\nabla^2 d^{(2)}_{({\rm t})i} & = 
&-\frac{4}{3}\nabla^2\beta^{(2)}_{({\rm t}),i}
-\chi^{(2)j}_{{\scriptscriptstyle \rm S}({\rm t})i,j}
-\frac{2\tau}{3} \left(\chi_{ij}^{(1)\top \prime}+\frac{4}{\tau}
\chi_{ij}^{(1)\top}\right)\varphi^{,j}
-\frac{2\tau^2}{3} \chi_{ij,k}^{(1)\top} \varphi^{,jk}\nonumber \\
&&
-\frac{\tau^2}{3}\chi_{ij}^{(1)\top} \nabla^2 \varphi^{,j}
-\frac{\tau^2}{9}\chi_{jk}^{(1)\top} \varphi^{~,jk}_{,i}
+\frac{2\tau^2}{9} \chi_{jk,i}^{(1)\top} \varphi^{,jk}
\nonumber.
\end{eqnarray}

The energy constraint (\ref{eq:ec}) can be used to replace 
$\chi^{(2),ij}_{{\scriptscriptstyle \rm S}({\rm t})\ ij}$
in term of ${\phi^{(2)}_{{\scriptscriptstyle \rm S}({\rm t})}}$ 
and products of first-order quantities.

On the other hand, we have from Eqs.\ (\ref{eq:psiP}) to 
(\ref{eq:chiP})
that the contributions to the perturbed metric that we are 
interested in are

\begin{eqnarray}
\label{eq:psipt}
\psi^{(2)}_{{\scriptscriptstyle \rm P}({\rm t})}&=&
\alpha^{(2)\prime}_{({\rm t})}+\frac{2}{\tau}\alpha^{(2)}_{({\rm t})}\;,
\\
\omega^{(2)}_{{\scriptscriptstyle \rm P}({\rm t})\ i}&=&
\frac{2\tau}{3} \chi_{ij}^{(1)\top}\varphi^{,j}
-\alpha^{(2)}_{({\rm t}),i} +\beta^{(2)\prime}_{({\rm t}),i} 
+d^{(2) \prime}_{({\rm t})i}\;,
\label{eq:omegapt}\\
\phi^{(2)}_{{\scriptscriptstyle \rm P}({\rm t})}& =
 & \phi^{(2)}_{{\scriptscriptstyle \rm S}({\rm t})}
-\frac{\tau^2}{9}\chi_{ij}^{(1)\top}\varphi^{,ij}
-\frac{2}{\tau}\alpha^{(2)}_{({\rm t})} 
-\frac{1}{3}\nabla^2\beta^{(2)}_{({\rm t})}\;,
\label{eq:phipt}\\
\chi^{(2)}_{{\scriptscriptstyle \rm P}({\rm t})\ ij} & = & 
\chi^{(2)}_{{\scriptscriptstyle \rm S}({\rm t})\ ij}
+\frac{2\tau}{3} \varphi\left(\chi_{ij}^{(1)\top \prime}
+\frac{4}{\tau}\chi_{ij}^{(1)\top}\right)
+\frac{\tau^2}{3}\left(\chi_{ij,k}^{(1)\top}\varphi^{,k}
+\chi_{ik}^{(1)\top}\varphi^{,k}_{,j}
+\chi_{jk}^{(1)\top}\varphi^{,k}_{~,i}\right)
  \nonumber \\
& &-\frac{2\tau^2}{9} \delta_{ij}\chi_{lk}^{(1)\top}\varphi^{,lk}
+2\left(d^{(2)}_{({\rm t})(i,j)} +
{\rm D}_{ij}\beta^{(2)}_{({\rm t})}\right)\; .
\label{eq:chipt}
\end{eqnarray} 

We can obtain expressions in terms of the synchronous 
gauge perturbed metric as follows:

\bigskip
\noindent
{\bf Lapse perturbation}
\bigskip

Replacing the expression for $\alpha^{(2)}_{({\rm t})}$ into Eq.\
(\ref{eq:psipt}) we obtain an expression in terms of 
$\phi^{(2)}_{{\scriptscriptstyle \rm S}({\rm t})}$, its derivatives 
and products of first-order quantities, that with the help of the
Raychaudhuri equation (\ref{eq:ray}) can be written as
\begin{eqnarray}
\nabla^2\nabla^2\psi^{(2)}_{{\scriptscriptstyle \rm P}({\rm t})}&=&
\frac{6}{\tau^2}\nabla^2\phi^{(2)}_{{\scriptscriptstyle \rm S}({\rm t})}
+\frac{6}{\tau}\chi_{ij}^{(1)\top \prime} \varphi^{,ij}
-3\nabla^2\chi_{ij}^{(1)\top} \varphi^{,ij}
+\frac{5}{8}\nabla^2(\chi^{(1)\top\prime\ ij}\chi_{ij}^{(1)\top \prime})
+\frac{1}{2\tau}\chi^{(1)\top\prime\ ij}\nabla^2\chi_{ij}^{(1)\top}
\nonumber\\
&&
-\frac{1}{4}\nabla^2\chi_{ij}^{(1)\top}\nabla^2\chi^{(1)\top\ ij}
-\frac{1}{\tau}\nabla^2\chi_{ij}^{(1)\top \prime}\chi^{(1)\top\ ij}
+\frac{1}{2}\chi^{(1)\top\ ij}\nabla^2\nabla^2\chi_{ij}^{(1)\top}
-\frac{1}{2\tau}\chi_{ij,k}^{(1)\top \prime}\chi^{(1)\top ij,k}
\nonumber\\
&&
+\frac{5}{4}\chi^{(1)\top\ ij,k}\nabla^2\chi_{ij,k}^{(1)\top}
-\frac{3}{2}\chi_{ij,k}^{(1)\top \prime}\chi^{(1)\top\prime\ kj,i}
+\frac{3}{\tau}\chi_{ij,k}^{(1)\top}\chi^{(1)\top\prime\ kj,i}
-\frac{3}{2}\chi^{(1)\top\ kj,i}\nabla^2\chi_{ij,k}^{(1)\top}
\nonumber\\
&&
+\frac{3}{\tau^2}\left(2\chi^{(1)\top\ ij}\nabla^2\chi_{ij}^{(1)\top}
+2\chi_{ij,k}^{(1)\top}\chi^{(1)\top\ ij,k}
-2\nabla^2\chi_{0\ ij}^{(1)\top}\chi_0^{(1)\top\ ij}
-2\chi_{0\ ij,k}^{(1)\top}\chi_0^{(1)\top\ ij,k}\right) \;.
\end{eqnarray}
\bigskip
\noindent
{\bf Shift perturbation}

>From Eq.\ (\ref{eq:omegapt}) and using the momentum constraint
(\ref{eq:mc}) and the Raychaudhuri equation (\ref{eq:ray}) we obtain
\begin{eqnarray}
\nabla^2\nabla^2\omega^{(2)}_{{\scriptscriptstyle \rm P}({\rm t})\ i}&=&
\nabla^2\left(-4  \chi_{ij}^{(1)\top \prime} \varphi^{,j}
-2 \chi_{ik,j}^{(1)\top \prime}\chi^{(1)\top\ jk}
+2 \chi_{jk,i}^{(1)\top \prime}\chi^{(1)\top\ jk}
+\chi_{jk}^{(1)\top \prime}\chi^{(1)\top\ jk}_{,i}\right)
\nonumber\\
&&+\left(4 \chi_{kj}^{(1)\top \prime} \varphi^{,kj}
-\chi_{kj}^{(1)\top \prime}\nabla^2\chi^{(1)\top\ kj}
-2 \chi^{(1)\top\ kj}\nabla^2 \chi_{kj}^{(1)\top \prime}
\right.
\nonumber\\
&&
\left.
-3\chi_{kj,l}^{(1)\top \prime}\chi^{(1)\top\ kj,l}
+2\chi_{kj,l}^{(1)\top \prime}\chi^{(1)\top\ lj,k}
\right)_{,i} \;. 
\end{eqnarray}
\bigskip
\noindent
{\bf Spatial metric, trace}

>From Eq.\ (\ref{eq:phipt}) and using the  Raychaudhuri equation (\ref{eq:ray})
we obtain 
\begin{eqnarray}
\nabla^2\nabla^2\phi^{(2)}_{{\scriptscriptstyle \rm P}({\rm t})}&=&
\frac{18}{\tau^2}\nabla^2\phi^{(2)}_{{\scriptscriptstyle \rm S}({\rm t})}
+\frac{1}{8}\nabla^2(\chi^{(1)\top\prime\ ij}\chi_{ij}^{(1)\top \prime})
-\frac{1}{2}\nabla^2\left(\chi_{ij}^{(1)\top}\nabla^2\chi^{(1)\top\ ij}
\right)
-\frac{3}{4}\chi_{ij,k}^{(1)\top}\nabla^2\chi^{(1)\top\ ij,k}
\nonumber\\
&&
-\frac{3}{4}\chi_{ij,kl}^{(1)\top}\chi^{(1)\top\ ij,kl}
+\frac{1}{2}\chi_{ij,k}^{(1)\top}\nabla^2\chi^{(1)\top\ kj,i}
+\frac{1}{2}\chi_{ij,kl}^{(1)\top}\chi^{(1)\top\ kj,il}
-\nabla^2(\chi_{ij}^{(1)\top} \varphi^{,ij})
-8 \varphi^{,ij}\nabla^2\chi_{ij}^{(1)\top}
\nonumber\\
&&
+\frac{1}{2\tau}\chi^{(1)\top\prime\ ij}\nabla^2\chi_{ij}^{(1)\top}
+\frac{6}{\tau}\chi_{ij}^{(1)\top \prime} \varphi^{,ij}
+\frac{12}{\tau^2}\chi_{ij}^{(1)\top \prime}\chi^{(1)\top\prime\ ij}
-\frac{1}{\tau}\chi^{(1)\top\ ij}\nabla^2\chi_{ij}^{(1)\top \prime}
\nonumber\\
&&
-\frac{1}{2\tau}\chi_{ij,k}^{(1)\top \prime}\chi^{(1)\top\ ij,k}
+\frac{3}{\tau}\chi_{ij,k}^{(1)\top}\chi^{(1)\top\prime\ kj,i}
+\frac{3}{\tau^2}\nabla^2\left(\chi_{ij}^{(1)\top}\chi^{(1)\top\ ij}
+\chi_0^{(1)\top\ ij}\chi_{0\ ij}^{(1)\top}\right)
\nonumber\\
&&
+\frac{24}{\tau^2}\chi_{ij}^{(1)\top}\nabla^2\chi^{(1)\top\ ij}
-\frac{48}{\tau^3}\chi_{ij}^{(1)\top \prime}\chi^{(1)\top\ ij}
-\frac{72}{\tau^4}\left(\chi_{ij}^{(1)\top}\chi^{(1)\top\ ij}
-\chi_{0\ ij}^{(1)\top}\chi_0^{(1)\top\ ij}\right) \;. 
\end{eqnarray}
\bigskip
\noindent
{\bf Spatial metric, traceless part}

Replacing from Eqs.\ (\ref{eq:beta2t}) the expressions for 
$\alpha^{(2)}_{({\rm t})}$ and $d^{(2)}_{({\rm t}) i}$ in
Eq.\ (\ref{eq:chipt}) we obtain
\begin{eqnarray}
\nabla^2\nabla^2\chi^{(2)}_{{\scriptscriptstyle \rm P}({\rm t})ij}&=&
\nabla^2\nabla^2\chi^{(2)}_{{\scriptscriptstyle \rm S}({\rm t})ij}
-2\nabla^2\chi^{(2),k}_{{\scriptscriptstyle \rm S}({\rm t})~k(i,j)}
+\frac{1}{2}\chi^{(2)kl}_{{\scriptscriptstyle \rm S}({\rm t})~,klij}
+\frac{1}{2}\delta_{ij}\nabla^2
\chi^{(2)kl}_{{\scriptscriptstyle \rm S}({\rm t})~,kl}
\nonumber\\
&&
-\nabla^2\left[\frac{4\tau}{3} \left(\varphi^{,k}
\left(\chi^{(1)\top\prime}_{k(i}+\frac{4}{\tau}\chi^{(1)\top}_{k(i}\right)
\right)_{,j)}
+\frac{4\tau^2}{3} \varphi^{,kl}\chi^{(1)\top}_{k(i,j)l}\right.
\nonumber\\
&&
+\frac{2\tau^2}{3} \nabla^2\varphi^{,l}\chi^{(1)\top}_{l(i,j)}
-\frac{4\tau^2}{9} \left(\chi^{(1)\top}_{lk,ij}\varphi^{,lk}
-\frac{1}{2}\chi^{(1)\top}_{lk}\varphi^{,lk}_{,ij}
+\frac{1}{2}\chi^{(1)\top}_{lk,(i}\varphi^{~~,lk}_{,j)}\right)
\nonumber\\
&&
-\frac{2\tau^2}{3} \nabla^2\chi^{(1)\top}_{k(i}\varphi^{~~,k}_{,j)}
-\nabla^2\left(\frac{2\tau}{3} \varphi\left(
\chi^{(1)\top\prime}_{ij}+\frac{4}{\tau}\chi^{(1)\top}_{ij}\right)
+\frac{\tau^2}{3} \varphi^{,k}\chi^{(1)\top}_{ij,k}\right)
\nonumber\\
&&\left.
-\frac{1}{2}\delta_{ij}\left(\frac{2\tau}{3} \varphi^{,kl}
(\chi^{(1)\top\prime}_{kl}+\frac{4}{\tau}\chi^{(1)\top}_{kl})
-\frac{\tau^2}{3} \varphi^{,klm}\chi^{(1)\top}_{kl,m}
-\frac{2\tau^2}{3} \varphi^{,kl}\nabla^2\chi^{(1)\top}_{kl}\right)
\right]
\nonumber\\
&&
+\frac{1}{2}\left[\frac{2\tau}{3} \varphi^{,kl}
(\chi^{(1)\top\prime}_{kl}+\frac{4}{\tau}\chi^{(1)\top}_{kl})
+\frac{5\tau^2}{9} \varphi^{,klm}\chi^{(1)\top}_{kl,m}
+\frac{4}{9}\tau^2\chi^{(1)\top}_{kl}\nabla^2\varphi^{,kl}
-\frac{2\tau^2}{9} \varphi^{,kl}\nabla^2\chi^{(1)\top}_{kl}
\right]_{,ij} \;.
\end{eqnarray}


\begin{references}
\bibitem{parameter}
For non linear perturbations the parametrization is not unique, see
Ref.\cite{bi:P1} for a brief discussion of this point; here we simply
assume that a parametrization has been given, and we work within it.
\bibitem{notation}
Note the difference  between Eq.\ (\ref{eq:expxmu}), where the
functions $\tilde{x}^\mu$ are given by the action of the
operator $\exp [\lambda\pounds_\xi]$ on the functions $x^\mu$, 
 and Eq.\ (\ref{eq:expZ}), where the components $\tilde{Z}^\mu$ of the
pull-back $\tilde{Z}$ are the $\mu$-components of the vector field given by
the action of $\exp [\lambda\pounds_\xi]$ on $Z$.
\end{references}
\end{document}